\documentclass[traditabstract, longauth]{aa}
\usepackage{graphicx}
\usepackage{amsmath,amsfonts,amssymb}
\usepackage{txfonts}
\usepackage[breaklinks,colorlinks,citecolor=blue,pdfa=true]{hyperref}
\usepackage{color}
\usepackage{fixltx2e}
\usepackage{natbib}
\usepackage{url}
\usepackage{multirow}
\usepackage{epsf}
\usepackage{epsfig}
\usepackage{ifthen}
\usepackage[T1]{fontenc}
\usepackage{lmodern}
\usepackage{ifxetex,ifluatex}
\usepackage{latexsym}

\usepackage{dblfloatfix}
\usepackage{morefloats}

\bibpunct{(}{)}{;}{a}{}{,} 

\newcommand{\nilc}{{\tt NILC}}
\newcommand{\sevem}{{\tt SEVEM}}
\newcommand{\smica}{{\tt SMICA}}
\newcommand{\commander}{{\tt Commander}}
\newcommand{\twodilc}{{\tt 2D-ILC}}

\newcommand{\gsim}{\; ^{>}_{\sim}\;}

\newcommand{\be}{\begin{equation}}
\newcommand{\ee}{\end{equation}}

\usepackage{multirow}

\def \der {{\rm d}}
\def \kms{{\rm km\,s^{-1}}}
\def \rms{{\it rms}}

\providecommand{\sorthelp}[1]{}

\def\setsymbol#1#2{\expandafter\def\csname #1\endcsname{#2}}
\def\getsymbol#1{\csname #1\endcsname}

\def\Planck{\textit{Planck}}

\newbox\tablebox    \newdimen\tablewidth
\def\leaderfil{\leaders\hbox to 5pt{\hss.\hss}\hfil}
\def\endPlancktable{\tablewidth=\columnwidth 
    $$\hss\copy\tablebox\hss$$
    \vskip-\lastskip\vskip -2pt}
\def\endPlancktablewide{\tablewidth=\textwidth 
    $$\hss\copy\tablebox\hss$$
    \vskip-\lastskip\vskip -2pt}
\def\tablenote#1 #2\par{\begingroup \parindent=0.8em
    \abovedisplayshortskip=0pt\belowdisplayshortskip=0pt
    \noindent
    $$\hss\vbox{\hsize\tablewidth \hangindent=\parindent \hangafter=1 \noindent
    \hbox to \parindent{$^#1$\hss}\strut#2\strut\par}\hss$$
    \endgroup}
\def\doubleline{\vskip 3pt\hrule \vskip 1.5pt \hrule \vskip 5pt}

\def\L2{\ifmmode L_2\else $L_2$\fi}

\def\DeltaT{\ifmmode \Delta T\else $\Delta T$\fi}
\def\deltat{\ifmmode \Delta t\else $\Delta t$\fi}
\def\fknee{\ifmmode f_{\rm knee}\else $f_{\rm knee}$\fi}
\def\Fmax{\ifmmode F_{\rm max}\else $F_{\rm max}$\fi}
\def\solar{\ifmmode{\rm M}_{\mathord\odot}\else${\rm M}_{\mathord\odot}$\fi}
\def\Msolar{\ifmmode{\rm M}_{\mathord\odot}\else${\rm M}_{\mathord\odot}$\fi}
\def\Lsolar{\ifmmode{\rm L}_{\mathord\odot}\else${\rm L}_{\mathord\odot}$\fi}
\def\inv{\ifmmode^{-1}\else$^{-1}$\fi}
\def\mo{\ifmmode^{-1}\else$^{-1}$\fi}
\def\sup#1{\ifmmode ^{\rm #1}\else $^{\rm #1}$\fi}
\def\expo#1{\ifmmode \times 10^{#1}\else $\times 10^{#1}$\fi}
\def\,{\thinspace}
\def\lsim{\mathrel{\raise .4ex\hbox{\rlap{$<$}\lower 1.2ex\hbox{$\sim$}}}}
\def\gsim{\mathrel{\raise .4ex\hbox{\rlap{$>$}\lower 1.2ex\hbox{$\sim$}}}}

\def\simprop{\mathrel{\raise .4ex\hbox{\rlap{$\propto$}\lower 1.2ex\hbox{$\sim$}}}}
\def\deg{\ifmmode^\circ\else$^\circ$\fi}
\def\pdeg{\ifmmode $\setbox0=\hbox{$^{\circ}$}\rlap{\hskip.11\wd0 .}$^{\circ}
          \else \setbox0=\hbox{$^{\circ}$}\rlap{\hskip.11\wd0 .}$^{\circ}$\fi}
\def\arcs{\ifmmode {^{\scriptstyle\prime\prime}}
          \else $^{\scriptstyle\prime\prime}$\fi}
\def\arcm{\ifmmode {^{\scriptstyle\prime}}
          \else $^{\scriptstyle\prime}$\fi}
\newdimen\sa  \newdimen\sb
\def\parcs{\sa=.07em \sb=.03em
     \ifmmode \hbox{\rlap{.}}^{\scriptstyle\prime\kern -\sb\prime}\hbox{\kern -\sa}
     \else \rlap{.}$^{\scriptstyle\prime\kern -\sb\prime}$\kern -\sa\fi}
\def\parcm{\sa=.08em \sb=.03em
     \ifmmode \hbox{\rlap{.}\kern\sa}^{\scriptstyle\prime}\hbox{\kern-\sb}
     \else \rlap{.}\kern\sa$^{\scriptstyle\prime}$\kern-\sb\fi}
\def\ra[#1 #2 #3.#4]{#1\sup{h}#2\sup{m}#3\sup{s}\llap.#4}
\def\dec[#1 #2 #3.#4]{#1\deg#2\arcm#3\arcs\llap.#4}
\def\deco[#1 #2 #3]{#1\deg#2\arcm#3\arcs}
\def\rra[#1 #2]{#1\sup{h}#2\sup{m}}

\def\dots{\relax\ifmmode \ldots\else $\ldots$\fi}
\def\WHzsr{\ifmmode $W\,Hz\mo\,sr\mo$\else W\,Hz\mo\,sr\mo\fi}
\def\mHz{\ifmmode $\,mHz$\else \,mHz\fi}
\def\GHz{\ifmmode $\,GHz$\else \,GHz\fi}
\def\mKs{\ifmmode $\,mK\,s$^{1/2}\else \,mK\,s$^{1/2}$\fi}
\def\muKs{\ifmmode \,\mu$K\,s$^{1/2}\else \,$\mu$K\,s$^{1/2}$\fi}
\def\muKRJs{\ifmmode \,\mu$K$_{\rm RJ}$\,s$^{1/2}\else \,$\mu$K$_{\rm RJ}$\,s$^{1/2}$\fi}
\def\muKHz{\ifmmode \,\mu$K\,Hz$^{-1/2}\else \,$\mu$K\,Hz$^{-1/2}$\fi}
\def\MJysr{\ifmmode \,$MJy\,sr\mo$\else \,MJy\,sr\mo\fi}
\def\MJysrmK{\ifmmode \,$MJy\,sr\mo$\,mK$_{\rm CMB}\mo\else \,MJy\,sr\mo\,mK$_{\rm CMB}\mo$\fi}
\def\microns{\ifmmode \,\mu$m$\else \,$\mu$m\fi}

\def\muK{\ifmmode \,\mu$K$\else \,$\mu$\hbox{K}\fi}
\def\microK{\ifmmode \,\mu$K$\else \,$\mu$\hbox{K}\fi}
\def\muW{\ifmmode \,\mu$W$\else \,$\mu$\hbox{W}\fi}
\def\kms{\ifmmode $\,km\,s$^{-1}\else \,km\,s$^{-1}$\fi}
\def\kmsMpc{\ifmmode $\,\kms\,Mpc\mo$\else \,\kms\,Mpc\mo\fi}
\providecommand{\sorthelp}[1]{}

\begin{document}

\author{
\author{\small
Planck Collaboration: N.~Aghanim\inst{48}
\and
Y.~Akrami\inst{50, 51}
\and
M.~Ashdown\inst{58, 4}
\and
J.~Aumont\inst{83}
\and
C.~Baccigalupi\inst{70}
\and
M.~Ballardini\inst{17, 35}
\and
A.~J.~Banday\inst{83, 7}
\and
R.~B.~Barreiro\inst{53}
\and
N.~Bartolo\inst{23, 54}
\and
S.~Basak\inst{75}
% ...
}
\institute{\small
APC, AstroParticule et Cosmologie, Universit\'{e} Paris Diderot, CNRS/IN2P3, CEA/lrfu, Observatoire de Paris, Sorbonne Paris Cit\'{e}, 10, rue Alice Domon et L\'{e}onie Duquet, 75205 Paris Cedex 13, France\goodbreak
\and
African Institute for Mathematical Sciences, 6-8 Melrose Road, Muizenberg, Cape Town, South Africa\goodbreak
\and
Agenzia Spaziale Italiana, Via del Politecnico snc, 00133, Roma, Italy\goodbreak
\and
Astrophysics Group, Cavendish Laboratory, University of Cambridge, J J Thomson Avenue, Cambridge CB3 0HE, U.K.\goodbreak
\and
Astrophysics \& Cosmology Research Unit, School of Mathematics, Statistics \& Computer Science, University of KwaZulu-Natal, Westville Campus, Private Bag X54001, Durban 4000, South Africa\goodbreak
\and
CITA, University of Toronto, 60 St. George St., Toronto, ON M5S 3H8, Canada\goodbreak
\and
CNRS, IRAP, 9 Av. colonel Roche, BP 44346, F-31028 Toulouse cedex 4, France\goodbreak
\and
California Institute of Technology, Pasadena, California, U.S.A.\goodbreak
\and
Computational Cosmology Center, Lawrence Berkeley National Laboratory, Berkeley, California, U.S.A.\goodbreak
\and
DTU Space, National Space Institute, Technical University of Denmark, Elektrovej 327, DK-2800 Kgs. Lyngby, Denmark\goodbreak
\and
D\'{e}partement de Physique Th\'{e}orique, Universit\'{e} de Gen\`{e}ve, 24, Quai E. Ansermet,1211 Gen\`{e}ve 4, Switzerland\goodbreak
\and
Departamento de Astrof\'{i}sica, Universidad de La Laguna (ULL), E-38206 La Laguna, Tenerife, Spain\goodbreak
\and
Departamento de F\'{\i}sica, Universidad de Oviedo, C/ Federico Garc\'{\i}a Lorca, 18 , Oviedo, Spain\goodbreak
\and
Department of Astrophysics/IMAPP, Radboud University, P.O. Box 9010, 6500 GL Nijmegen, The Netherlands\goodbreak
\and
Department of Mathematics, University of Stellenbosch, Stellenbosch 7602, South Africa\goodbreak
\and
Department of Physics \& Astronomy, University of British Columbia, 6224 Agricultural Road, Vancouver, British Columbia, Canada\goodbreak
\and
Department of Physics \& Astronomy, University of the Western Cape, Cape Town 7535, South Africa\goodbreak
\and
Department of Physics and Astronomy, University of Sussex, Brighton BN1 9QH, U.K.\goodbreak
\and
Department of Physics, Gustaf H\"{a}llstr\"{o}min katu 2a, University of Helsinki, Helsinki, Finland\goodbreak
\and
Department of Physics, Princeton University, Princeton, New Jersey, U.S.A.\goodbreak
\and
Department of Physics, University of California, Santa Barbara, California, U.S.A.\goodbreak
\and
Department of Physics, University of Illinois at Urbana-Champaign, 1110 West Green Street, Urbana, Illinois, U.S.A.\goodbreak
\and
Dipartimento di Fisica e Astronomia G. Galilei, Universit\`{a} degli Studi di Padova, via Marzolo 8, 35131 Padova, Italy\goodbreak
\and
Dipartimento di Fisica e Scienze della Terra, Universit\`{a} di Ferrara, Via Saragat 1, 44122 Ferrara, Italy\goodbreak
\and
Dipartimento di Fisica, Universit\`{a} La Sapienza, P. le A. Moro 2, Roma, Italy\goodbreak
\and
Dipartimento di Fisica, Universit\`{a} degli Studi di Milano, Via Celoria, 16, Milano, Italy\goodbreak
\and
Dipartimento di Fisica, Universit\`{a} degli Studi di Trieste, via A. Valerio 2, Trieste, Italy\goodbreak
\and
Dipartimento di Fisica, Universit\`{a} di Roma Tor Vergata, Via della Ricerca Scientifica, 1, Roma, Italy\goodbreak
\and
European Space Agency, ESAC, Planck Science Office, Camino bajo del Castillo, s/n, Urbanizaci\'{o}n Villafranca del Castillo, Villanueva de la Ca\~{n}ada, Madrid, Spain\goodbreak
\and
European Space Agency, ESTEC, Keplerlaan 1, 2201 AZ Noordwijk, The Netherlands\goodbreak
\and
Gran Sasso Science Institute, INFN, viale F. Crispi 7, 67100 L'Aquila, Italy\goodbreak
\and
HGSFP and University of Heidelberg, Theoretical Physics Department, Philosophenweg 16, 69120, Heidelberg, Germany\goodbreak
\and
Haverford College Astronomy Department, 370 Lancaster Avenue, Haverford, Pennsylvania, U.S.A.\goodbreak
\and
Helsinki Institute of Physics, Gustaf H\"{a}llstr\"{o}min katu 2, University of Helsinki, Helsinki, Finland\goodbreak
\and
INAF - OAS Bologna, Istituto Nazionale di Astrofisica - Osservatorio di Astrofisica e Scienza dello Spazio di Bologna, Area della Ricerca del CNR, Via Gobetti 101, 40129, Bologna, Italy\goodbreak
\and
INAF - Osservatorio Astronomico di Padova, Vicolo dell'Osservatorio 5, Padova, Italy\goodbreak
\and
INAF - Osservatorio Astronomico di Trieste, Via G.B. Tiepolo 11, Trieste, Italy\goodbreak
\and
INAF, Istituto di Radioastronomia, Via Piero Gobetti 101, I-40129 Bologna, Italy\goodbreak
\and
INAF/IASF Milano, Via E. Bassini 15, Milano, Italy\goodbreak
\and
INFN - CNAF, viale Berti Pichat 6/2, 40127 Bologna, Italy\goodbreak
\and
INFN, Sezione di Bologna, viale Berti Pichat 6/2, 40127 Bologna, Italy\goodbreak
\and
INFN, Sezione di Ferrara, Via Saragat 1, 44122 Ferrara, Italy\goodbreak
\and
INFN, Sezione di Milano, Via Celoria 16, Milano, Italy\goodbreak
\and
INFN, Sezione di Roma 1, Universit\`{a} di Roma Sapienza, Piazzale Aldo Moro 2, 00185, Roma, Italy\goodbreak
\and
INFN, Sezione di Roma 2, Universit\`{a} di Roma Tor Vergata, Via della Ricerca Scientifica, 1, Roma, Italy\goodbreak
\and
INFN/National Institute for Nuclear Physics, Via Valerio 2, I-34127 Trieste, Italy\goodbreak
\and
Imperial College London, Astrophysics group, Blackett Laboratory, Prince Consort Road, London, SW7 2AZ, U.K.\goodbreak
\and
Institut d'Astrophysique Spatiale, CNRS, Univ. Paris-Sud, Universit\'{e} Paris-Saclay, B\^{a}t. 121, 91405 Orsay cedex, France\goodbreak
\and
Institut d'Astrophysique de Paris, CNRS (UMR7095), 98 bis Boulevard Arago, F-75014, Paris, France\goodbreak
\and
Institute Lorentz, Leiden University, PO Box 9506, Leiden 2300 RA, The Netherlands\goodbreak
\and
Institute of Theoretical Astrophysics, University of Oslo, Blindern, Oslo, Norway\goodbreak
\and
Instituto de Astrof\'{\i}sica de Canarias, C/V\'{\i}a L\'{a}ctea s/n, La Laguna, Tenerife, Spain\goodbreak
\and
Instituto de F\'{\i}sica de Cantabria (CSIC-Universidad de Cantabria), Avda. de los Castros s/n, Santander, Spain\goodbreak
\and
Istituto Nazionale di Fisica Nucleare, Sezione di Padova, via Marzolo 8, I-35131 Padova, Italy\goodbreak
\and
Jet Propulsion Laboratory, California Institute of Technology, 4800 Oak Grove Drive, Pasadena, California, U.S.A.\goodbreak
\and
Jodrell Bank Centre for Astrophysics, Alan Turing Building, School of Physics and Astronomy, The University of Manchester, Oxford Road, Manchester, M13 9PL, U.K.\goodbreak
\and
Kavli Institute for Cosmological Physics, University of Chicago, Chicago, IL 60637, USA\goodbreak
\and
Kavli Institute for Cosmology Cambridge, Madingley Road, Cambridge, CB3 0HA, U.K.\goodbreak
\and
LAL, Universit\'{e} Paris-Sud, CNRS/IN2P3, Orsay, France\goodbreak
\and
LERMA, CNRS, Observatoire de Paris, 61 Avenue de l'Observatoire, Paris, France\goodbreak
\and
Laboratoire de Physique Subatomique et Cosmologie, Universit\'{e} Grenoble-Alpes, CNRS/IN2P3, 53, rue des Martyrs, 38026 Grenoble Cedex, France\goodbreak
\and
Laboratoire de Physique Th\'{e}orique, Universit\'{e} Paris-Sud 11 \& CNRS, B\^{a}timent 210, 91405 Orsay, France\goodbreak
\and
Lawrence Berkeley National Laboratory, Berkeley, California, U.S.A.\goodbreak
\and
Low Temperature Laboratory, Department of Applied Physics, Aalto University, Espoo, FI-00076 AALTO, Finland\goodbreak
\and
Max-Planck-Institut f\"{u}r Astrophysik, Karl-Schwarzschild-Str. 1, 85741 Garching, Germany\goodbreak
\and
Mullard Space Science Laboratory, University College London, Surrey RH5 6NT, U.K.\goodbreak
\and
NAOC-UKZN Computational Astrophysics Centre (NUCAC), University of KwaZulu-Natal, Durban 4000, South Africa\goodbreak
\and
Nicolaus Copernicus Astronomical Center, Polish Academy of Sciences, Bartycka 18, 00-716 Warsaw, Poland\goodbreak
\and
Nordita (Nordic Institute for Theoretical Physics), Roslagstullsbacken 23, SE-106 91 Stockholm, Sweden\goodbreak
\and
SISSA, Astrophysics Sector, via Bonomea 265, 34136, Trieste, Italy\goodbreak
\and
San Diego Supercomputer Center, University of California, San Diego, 9500 Gilman Drive, La Jolla, CA 92093, USA\goodbreak
\and
School of Chemistry and Physics, University of KwaZulu-Natal, Westville Campus, Private Bag X54001, Durban, 4000, South Africa\goodbreak
\and
School of Physics and Astronomy, Cardiff University, Queens Buildings, The Parade, Cardiff, CF24 3AA, U.K.\goodbreak
\and
School of Physics and Astronomy, Sun Yat-Sen University, 135 Xingang Xi Road, Guangzhou, China\goodbreak
\and
School of Physics, Indian Institute of Science Education and Research Thiruvananthapuram, Maruthamala PO, Vithura, Thiruvananthapuram 695551, Kerala, India\goodbreak
\and
Simon Fraser University, Department of Physics, 8888 University Drive, Burnaby BC, Canada\goodbreak
\and
Sorbonne Universit\'{e}-UPMC, UMR7095, Institut d'Astrophysique de Paris, 98 bis Boulevard Arago, F-75014, Paris, France\goodbreak
\and
Space Research Institute (IKI), Russian Academy of Sciences, Profsoyuznaya Str, 84/32, Moscow, 117997, Russia\goodbreak
\and
Space Science Data Center - Agenzia Spaziale Italiana, Via del Politecnico snc, 00133, Roma, Italy\goodbreak
\and
Space Sciences Laboratory, University of California, Berkeley, California, U.S.A.\goodbreak
\and
The Oskar Klein Centre for Cosmoparticle Physics, Department of Physics, Stockholm University, AlbaNova, SE-106 91 Stockholm, Sweden\goodbreak
\and
UPMC Univ Paris 06, UMR7095, 98 bis Boulevard Arago, F-75014, Paris, France\goodbreak
\and
Universit\'{e} de Toulouse, UPS-OMP, IRAP, F-31028 Toulouse cedex 4, France\goodbreak
% ...
}
}
\title{\Planck\ intermediate results. LIII. Detection of velocity dispersion
 from the kinetic Sunyaev-Zeldovich effect}
\authorrunning{Planck Collaboration}
\titlerunning{Velocity dispersion from the kSZ effect}

\date{\today}

\abstract{Using the \Planck\ full-mission data, we present a detection
of the temperature (and therefore velocity) dispersion due to the kinetic
Sunyaev-Zeldovich (kSZ) effect from clusters of galaxies. To suppress the
primary CMB and instrumental noise we derive a matched filter and then convolve
it with the \Planck\ foreground-cleaned ``{\tt 2D-ILC\,}'' maps. By using
the Meta Catalogue of X-ray detected Clusters of galaxies (MCXC),
we determine the normalized \rms\ dispersion
of the temperature fluctuations at the positions of clusters, finding that this
shows excess variance compared with the noise expectation. We then build an
unbiased statistical estimator of the signal, determining that the normalized
mean temperature dispersion of $1526$ clusters
is $\langle \left(\Delta T/T \right)^{2} \rangle = (1.64 \pm 0.48)
\times 10^{-11}$.
However, comparison with analytic calculations and simulations suggest that
around $0.7\,\sigma$ of this result is due to cluster lensing rather than the
kSZ effect. By correcting this, the temperature dispersion is measured to be
$\langle \left(\Delta T/T \right)^{2} \rangle = (1.35 \pm 0.48)
\times 10^{-11}$, which gives a detection at the $2.8\,\sigma$ level. We
further convert uniform-weight temperature
dispersion into a measurement of the line-of-sight
velocity dispersion, by using estimates of the optical depth of each cluster
(which introduces additional uncertainty into the estimate).
We find that the velocity dispersion is
$\langle v^{2} \rangle =(123\,000 \pm 71\,000)\,(\kms)^{2}$,
which is consistent with findings from other
large-scale structure studies, and provides direct evidence of statistical
homogeneity on scales of $600\,h^{-1}{\rm Mpc}$. Our study shows the promise of
using cross-correlations of the kSZ effect with large-scale structure in order
to constrain the growth of structure.}



\maketitle

\section{Introduction}
\label{sec:intro}

The kinetic Sunyaev-Zeldovich (hereafter kSZ; \citealt{kSZ0,kSZ}) effect
describes the temperature anisotropy of the cosmic microwave background (CMB)
radiation due to inverse Compton scattering of CMB photons off a moving cloud
of electrons.  The effect can be written as
\begin{eqnarray}
  \frac{\Delta T}{T}(\vec{\hat{r}})=-\frac{\sigma_{\rm T}}{c} \int n_{\rm e}
  \left(\vec{v}\cdot \vec{\hat{r}} \right) \der l ,\label{eq:kSZ1}
\end{eqnarray}
where $\sigma_{\rm T}$ is the Thomson cross-section, $n_{\rm e}$ is the
electron density, $\vec{v}\cdot \vec{\hat{r}}$ is the velocity along the
line-of-sight, and $\der l$ is the path length in the radial direction. By
adopting a so-called ``pairwise momentum estimator,'' i.e., using the weights
that quantify the difference in temperature between pairs of galaxies, the
effect was first detected by \cite{Handetal2012} using CMB maps from the
Atacama Cosmology Telescope (ACT). The detection of the kSZ effect has been
further solidified using the same pairwise momentum estimator with other CMB
data, including {\it Wilkinson Microwave Anisotropy Probe} ({\it WMAP}) 9-year W-band data, and \Planck's four foreground-cleaned
maps \citep{planck2015-XXXVII}, and again more recently using a Fourier space
analysis \citep{Sugiyama2017}. These measurements represent detections
at the $2$--$3\,\sigma$
level. In addition, \citet{Bernardis17} applied the pairwise momentum estimator to the ACT data and 50\,000 bright galaxies from BOSS DR11 catalogue, and obtained $3.6$--$4.1\,\sigma$ C.L. detection. By using the same estimator to the South Pole Telescope (SPT) data and Dark Energy Survey (DES) data, \citet{Soergel16} achieved the detection of kSZ signal $4.2\,\sigma$ C.L. More recently \citet{LiMa17} detected the pairwise kSZ signal for BOSS DR13 low mass group ($M_{\rm h} \lesssim 10^{12}\,h^{-1}{\rm M}_{\odot}$) by using the {\tt 2D-ILC} map (see Section~\ref{sec:2DILC}). Besides the pairwise momentum estimator, in \citet{planck2015-XXXVII} the kSZ temperature map
($\delta T$) was estimated from \Planck\ full-mission data and cross-correlated
with the reconstructed linear velocity field data ($\vec{v}\cdot
\vec{\hat{n}}$) from the Sloan Digital Sky Survey (SDSS-DR7)
to compute the correlation function
$\langle \Delta T (\vec{v}\cdot \vec{\hat{n}}) \rangle$.
For this cross-correlation, $3.0$--$3.2\,\sigma$
detections were found for the foreground-cleaned \Planck\ maps, and
$3.8\,\sigma$ for the \Planck\ 217-GHz map. Following
\citet{planck2015-XXXVII}, \citet{Schaan16} detected the aggregated signal of
the kSZ effect at $3.3\,\sigma$ C.L. by cross-correlating the velocity field from
BOSS samples with the kSZ map produced from ACT observations. More recently,
\citet{Hill16} cross-correlated the squared kSZ fields from {\it WMAP} and \Planck,
and the projected galaxy overdensity from the Wide-field Infrared Survey
Explorer (WISE), which led to a $3.8\,\sigma$ detection. With advanced ACTPol
and a future Stage-IV CMB experiment, the signal-to-noise (S/N) ratio of the
kSZ squared field and projected density field could reach $120$ and $150$,
respectively \citep{Ferraro16}, although these authors also cautioned that the
results should be corrected for a bias due to lensing.
These previous attempts to make various kinds
of kSZ measurement are summarized in Table~\ref{tab:history}.

There has been a lot of previous work investigating how to use kSZ measurements
to determine the peculiar velocity field. This idea was first proposed
by \cite{haehnelt95}, suggesting that on small angular scales the peculiar
velocities of clusters could be inferred from CMB observations. \cite{nabila01}
estimated the potential uncertainty of the kSZ measurements due to
contamination by the primary CMB and thermal Sunyaev-Zeldovich (hereafter tSZ)
effect. In \cite{holzapfeletal97}, the peculiar velocities of two distant
galaxy clusters, namely Abell 2163 ($z=0.201$) and Abell 1689 ($z=0.181$), were
estimated through millimetre-wavelength observations (the SZ Infrared
Experiment, SuZIE). Furthermore, \cite{bensonetal03} estimated the bulk flow
using six galaxy clusters at $z > 0.2$ from the SuZIE II experiment in three
frequency bands between 150 and 350\,GHz, constraining the bulk flow to be
$<1410 \kms$ at 95\,\% CL. In addition, \citet{kashlinsky00} and
\citet{kashlinsky08,kashlinsky09} estimated peculiar velocities on large scales
and claimed a ``dark flow'' ($\ga1000\,\kms$) on Gpc scales.
However, by combining
galaxy cluster catalogues with \Planck\ nominal mission foreground-cleaned
maps, \cite{planck2013-XIII} constrained the cluster velocity monopole to be
$72\pm 60 \kms$ and the dipole (bulk flow) to be $<254 \kms$ (95\,\% CL) in the
CMB rest frame. This indicates that the Universe is largely homogeneous on Gpc
scales, consistent with the standard $\Lambda$ cold dark matter ($\Lambda$CDM)
scenario with adiabatic initial conditions.\footnote{Although in principle one
could still have an isocurvature perturbation on very large scales
\citep{Turner91,Ma-Gordon11}.}

This work represents the third contribution of the \Planck\footnote{\Planck\
({\tt http://www.esa.int/Planck}) is a project of the European Space Agency
(ESA) with instruments provided by two scientific consortia funded by ESA
member states and led by Principal Investigators from France and Italy,
telescope reflectors provided through a collaboration between ESA and a
scientific consortium led and funded by Denmark, and additional contributions
from NASA (USA).} Collaboration to the study of the kSZ effect. In
\citet{planck2013-XIII} we focused on constraining the monopole and dipole of
the peculiar velocity field, which gives constraints on the large-scale
inhomogeneity of the Universe. In the second paper, \citet{planck2015-XXXVII},
we calculated the pairwise momentum of the kSZ effect and cross-correlated this
with the reconstructed peculiar velocity field $\langle \Delta T
(\vec{v}\cdot\vec{\hat{n}}) \rangle$, obtaining direct evidence of unbound gas
outside the virial radii of the clusters. A follow-up paper modelled these
results to reconstruct the baryon fraction and suggested that this unbound gas
corresponds to {\it all\/} baryons surrounding the galaxies \citep{chm15}. Even
though the large-scale bulk flow and monopole flow were not detected
in \cite{planck2013-XIII}, the small-scale velocity dispersion in the nearby
Universe, determined by the local gravitational potential field, might still be
measurable.
This is because the velocity of each galaxy comprises two components,
namely the bulk flow components, which reflect the large-scale perturbations,
and a small-scale velocity dispersion component, which reflects perturbations
due to the local gravitational potential \citep[see, e.g.,][]{watkinsetal09,
feldmanetal10, mascott12, mascott14}. Therefore, although the bulk flow of the
galaxy clusters is constrained to be less than $254\kms$, the total velocity
dispersion can still be large enough to be detected. With that motivation, in
this paper we will look at a different aspect than in the previous two papers,
namely focusing on 1-point statistics of \Planck\ data to constrain the
temperature and velocity dispersion due to the kSZ effect. This topic is
relevant for large-scale structure, since the velocity dispersion that
we are trying to measure can be used as a sensitive test for galaxy formation
models \citep{Ostriker80, Davies83, Kormendy96, Kormendy01, MacMillan06} and
moreover, a numerical value for the small-scale dispersion often has to be
assumed in studies of large-scale flows \citep[e.g.,][]{Ma-Gordon11,Turnbull12}.
Providing such a statistical test through \Planck's full-mission
foreground-cleaned maps is the main aim of the present paper.

This paper is organized as follows. In Sect.~\ref{sec:data}, we describe the
\Planck\ CMB data and the X-ray catalogue of detected clusters of galaxies. In
Sect.~\ref{sec:methodology}, we discuss the filter that we develop to convolve
the observational map, and the statistical methodology that we use for
searching for the kSZ temperature-dispersion signal. Then we present the
results of our search along with relevant statistical tests. In
Sect.~\ref{sec:velocity}, we discuss the astrophysical implications of our
result, the conclusions being presented in the last section. Throughout
this work, we adopt a spatially flat, $\Lambda$CDM cosmology model, with the
best-fit cosmological parameters given by \cite{planck2014-a15}: $\Omega_{\rm
m}=0.309$; $\Omega_{\Lambda} = 0.691$; $n_{\rm s} = 0.9608$; $\sigma_{8} =
0.809$; and $h = 0.68$, where the Hubble constant is $H_{0} = 100 h \kms\,{\rm
Mpc}^{-1}$.

\begin{table*}[htb!]
\begingroup
\newdimen\tblskip \tblskip=5pt
\caption{A list of the recent measurements of the kinetic Sunyaev-Zeldovich
effect with cross-correlations of various tracers of large-scale structure.
``Spec-$z$'' and ``photo-$z$'' mean galaxy surveys with spectroscopic or
photometric redshift data, respectively. ``DES'' stands for Dark Energy Survey.
}
\label{tab:history}
\vskip -3mm
\setbox\tablebox=\vbox{
   \newdimen\digitwidth
   \setbox0=\hbox{\rm 0}
   \digitwidth=\wd0
   \catcode`*=\active
   \def*{\kern\digitwidth}
   \newdimen\signwidth
   \setbox0=\hbox{+}
   \signwidth=\wd0
   \catcode`!=\active
   \def!{\kern\signwidth}
\halign{\tabskip0em#\hfil\tabskip 1em&
 #\hfil\tabskip 1em&
 #\hfil\tabskip 1em&
 #\hfil\tabskip 1em&
 #\hfil\tabskip 1em&
 #\hfil\tabskip 0pt\cr
\noalign{\doubleline}
\noalign{\vskip -1pt}
Method& Reference& kSZ data& Tracer type& Tracer data& Significance\cr
\noalign{\vskip 3pt\hrule\vskip 5pt}
Pairwise& \citet{Handetal2012}$^{\rm a}$& ACT& Galaxies (spec-$z$)& BOSS III/DR9& $2.9\,\sigma$\cr
temperature& \citet{planck2015-XXXVII}& \Planck& Galaxies (spec-$z$)& SDSS/DR7& $1.8$--$2.5\,\sigma$\cr
difference& \citet{chm15}& {\it WMAP}& Galaxies (spec-$z$)& SDSS/DR7& $3.3\,\sigma$\cr
& \citet{Soergel16}& SPT& Clusters (photo-$z$)& $1$-yr DES& $4.2\,\sigma$\cr
& \citet{Bernardis17}& ACT& Galaxies (spec-$z$)& BOSS/DR11& $3.6$--$4.1\,\sigma$\cr
& \citet{Sugiyama2017}$^{\rm b}$& \Planck& Galaxies (spec-$z$)& BOSS/DR12& $2.45\,\sigma$\cr
& \citet{LiMa17}$^{\rm b}$& \Planck& Galaxies (spec-$z$)& BOSS/DR12& $1.65\,\sigma$\cr
\noalign{\vskip 3pt\hrule\vskip 5pt}
${\rm kSZ}\times v_{\rm pec}$& \citet{planck2015-XXXVII}$^{\rm c}$& \Planck& Galaxy velocities& SDSS/DR7& $3.0$--$3.7\,\sigma$\cr
& \citet{Schaan16}$^{\rm c}$& ACT& Galaxy velocities& BOSS/DR10& $2.9\,\sigma$, $3.3\,\sigma$\cr
\noalign{\vskip 3pt\hrule\vskip 5pt}
${\rm kSZ}^2\times {\rm projected}$& \citet{Hill16},& {\it Planck},& Projected& WISE& $3.8$--$4.5\,\sigma$\cr
density field& \citet{Ferraro16}$^{\rm d}$& {\it WMAP} & overdensities& catalogue& \cr
\noalign{\vskip 3pt\hrule\vskip 5pt}
kSZ dispersion& This work& \Planck& Clusters& MCXC& $2.8\,\sigma$\cr
\noalign{\vskip 4pt\hrule\vskip 5pt}}}
\endPlancktablewide
\tablenote{{\rm a}} A $p$-value of $0.002$ is quoted in the original paper,
 which we convert to signal-to-noise ratio by using Eq.~(\ref{eq:SN-P}). \par
\tablenote{{\rm b}} The differences between \citet{Sugiyama2017}
and \citet{LiMa17} are that the former used a Fourier-space analysis and a
density-weighted pairwise estimator, while the latter used a real-space analysis
and a uniform-weighting pairwise momentum estimator. \par
\tablenote{{\rm c}} Galaxy peculiar velocities form SDSS/DR7 in
\citet{planck2015-XXXVII} and BOSS/DR10 in \citet{Schaan16} are obtained by
reconstructing the linear peculiar velocity field from the density field. \par
\tablenote{{\rm d}} Here ``kSZ$^2$'' means the squared kSZ field. \par
\endgroup
\end{table*}

\section{Data description}
\label{sec:data}

\subsection{\Planck\ maps}
\label{sec:planck}

\subsubsection{Maps from the Planck Legacy Archive}

In this work we use the publicly released \Planck\ 2015 data.\footnote{From the
Planck Legacy Archive, \url{http://pla.esac.esa.int}\,.}
The kSZ effect gives rise to frequency-independent temperature fluctuations
that are a source of secondary anisotropies. The kSZ effect should therefore
be present
in all CMB foreground-cleaned products. Here we investigate the four \Planck\
2015 foreground-cleaned maps, namely the \commander, \nilc, \sevem, and \smica\
maps.  These are the outputs of four different component-separation algorithms
\citep{planck2015-XXXVII} and have a resolution of $\theta_{\rm
FWHM}=5\,$arcmin. \smica\ uses a spectral-matching approach, \sevem\ adopts a
template-fitting method to minimize the foregrounds, \nilc\ is the result of an
internal linear combination approach, and \commander\ uses a parametric,
pixel-based Monte Carlo Markov chain technique to project out
foregrounds \citep{planck2013-p06, planck2014-a11, planck2014-a12}. All of
these maps are produced with the intention of minimizing the foreground
contribution, but there could nevertheless be some residual contamination from
the tSZ effect, as well as other foregrounds \citep[e.g., the Galactic kSZ
effect, see][]{waelkens2008}.  We use the {\tt HEALPix}
package \citep{gorski2005} to visualize and mainpulate the maps.

\subsubsection{The {\tt 2D-ILC\,} map}
\label{sec:2DILC}

\begin{figure}
\includegraphics[width=\columnwidth]{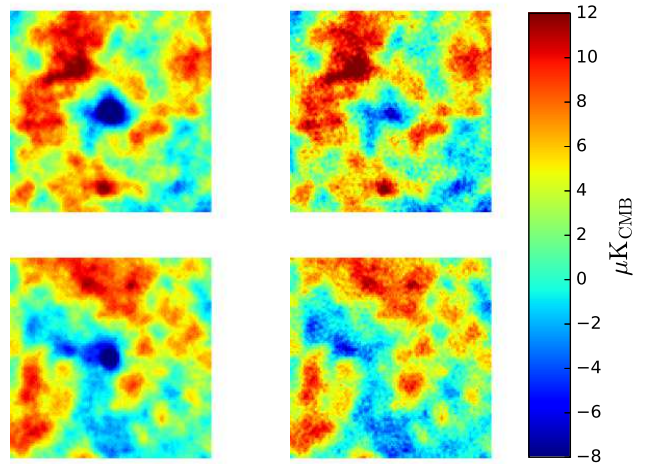}
\caption[fig:stacking]{\emph{Top left:} Stack of the
{\tt NILC} CMB map in the directions of \Planck\ SZ (PSZ) galaxy clusters.
\emph{Top right:} Stack of the {\tt 2D-ILC} CMB map
in the direction of PSZ galaxy clusters.  This sample provides a very stringent
test of the tSZ leakage, since the PSZ positions are the known places on the
sky with detectable SZ signal.  The stacked {\tt NILC} CMB map clearly shows
an excess in the centre, which is due to residual
contamination from the tSZ effect, while the {\tt 2D-ILC} CMB map has a
signature in the centre that is consistent with the strength of other features
in the stacked image.  \emph{Bottom left:} Stack of the
{\tt NILC} CMB map in the directions of MCXC clusters.  \emph{Bottom right:}
Stack of the {\tt 2D-ILC} CMB map
in the direction of $1526$ MCXC clusters (see Sect.~\ref{sec:MCXC} for the detail of the catalogue).  For a different set of sky positions, the
results are broadly consistent with those for the PSZ clusters.  All these
maps are $3^\circ \times 3^\circ$ in size, and use the same colour scale.}
\label{fig:stacking}
\end{figure}

\begin{figure}
\centering
\includegraphics[width=\columnwidth]{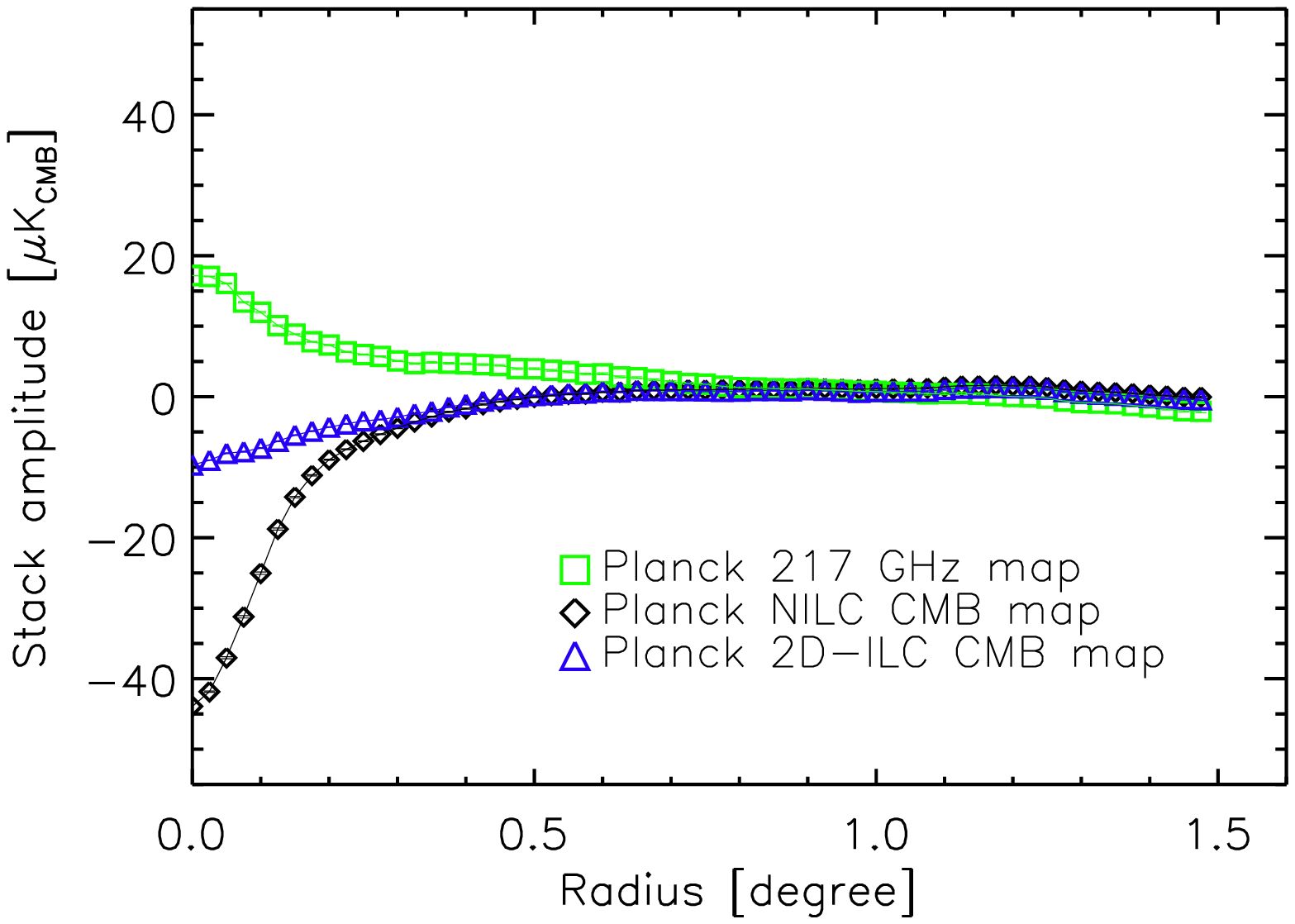}\\
\includegraphics[width=\columnwidth]{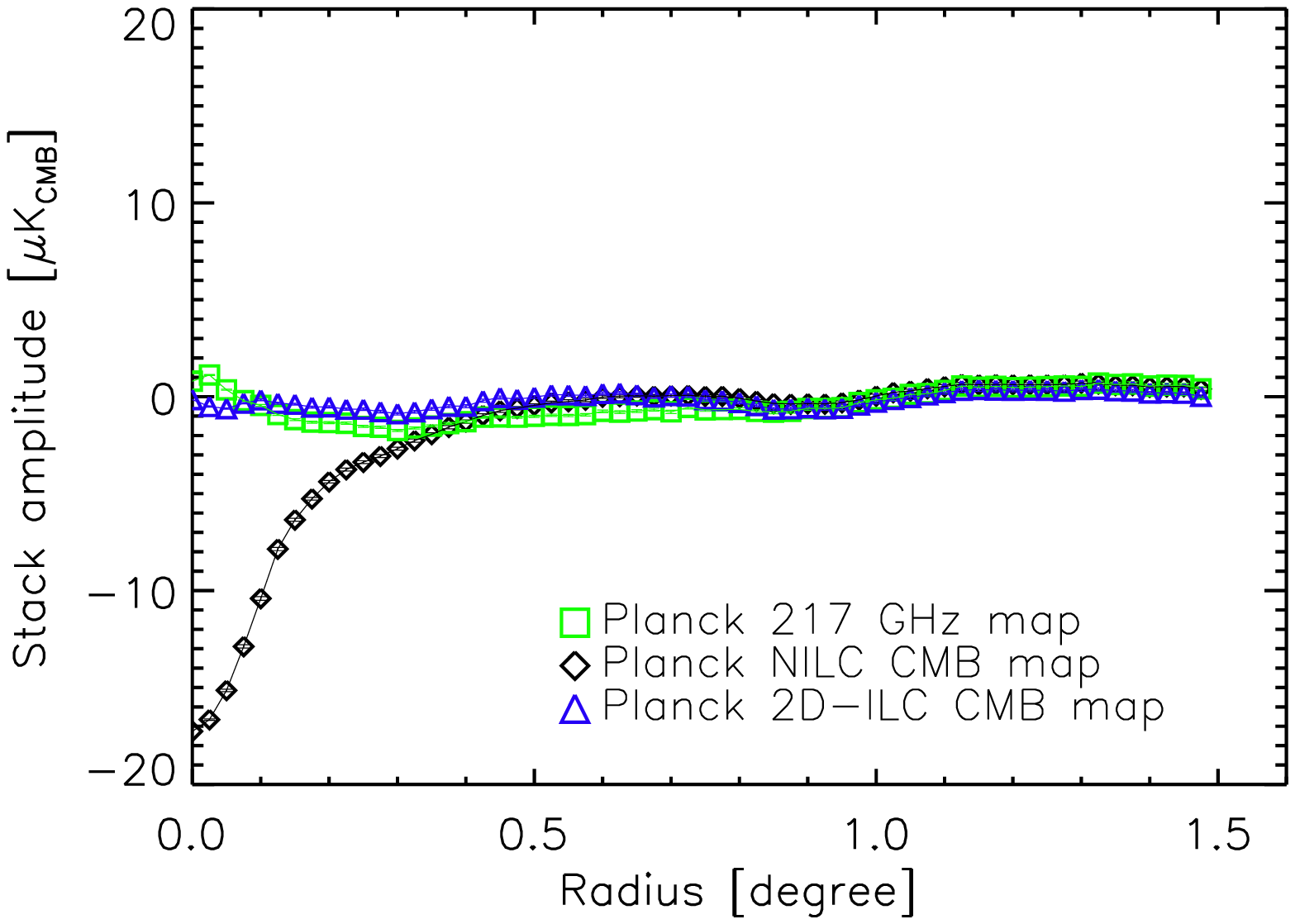}
\caption[fig:profiles]{Profiles for stacked patches (see
Fig.~\ref{fig:stacking}) of the {\tt NILC} CMB map (black diamonds) and the
{\tt 2D-ILC} CMB map (blue triangles) at the positions of PSZ clusters (\emph{top panel}) and MCXC clusters (\emph{bottom panel}).
The profile of the stacked \Planck\
217-GHz map is also shown as a reference (green squares). The central deficit
in the flux profile of the stacked {\tt NILC} CMB map (black diamonds) is due
to residual tSZ contamination.}
\label{fig:profiles}
\end{figure}

\begin{figure*}
\centering
\centerline{
\includegraphics[width=9.cm]{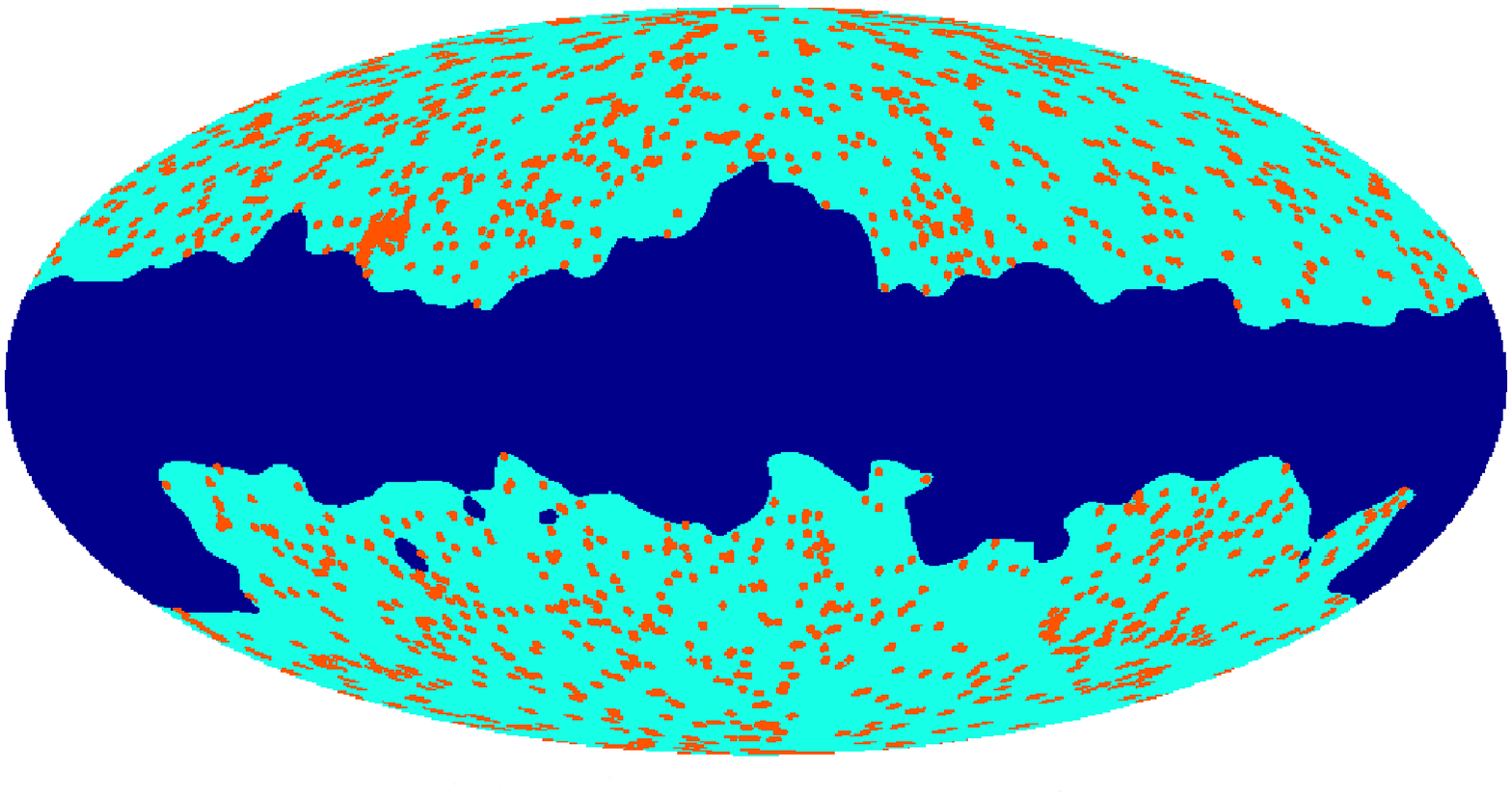}
\includegraphics[width=8.cm]{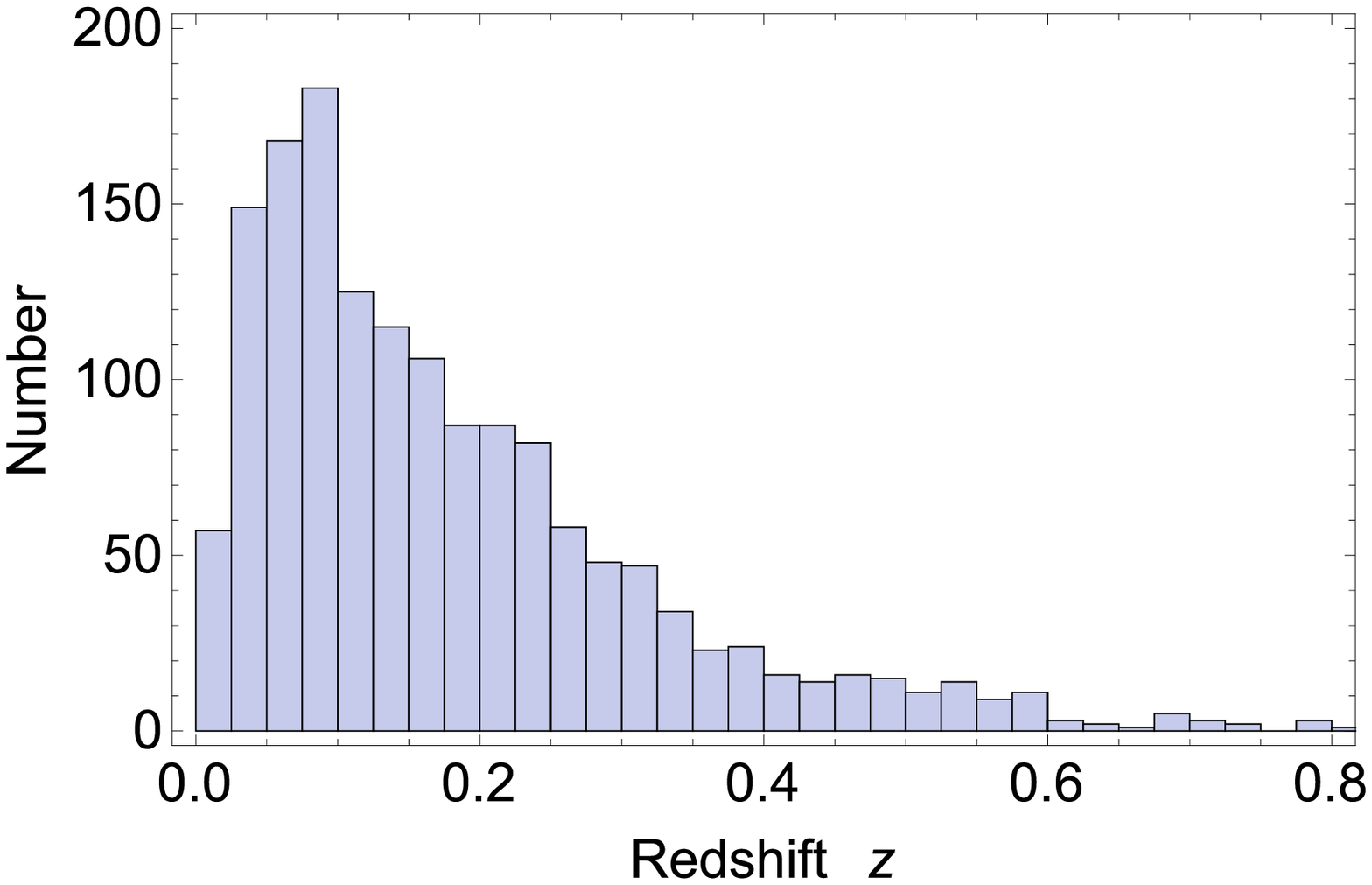}}
\caption[fig:cluster]{{\it Left}: Full-sky distribution of $1526$ MCXC X-ray
clusters \citep{pif,planck2013-XIII} in Galactic coordinates. The dark blue
area is the masked region, and the clusters are shown in orange. {\it Right}:
Redshift histogram of $1526$ X-ray clusters, with bin width $\Delta z=0.025$.}
\label{fig:cluster}
\end{figure*}

The {\tt 2D-ILC} \Planck\ CMB map has the additional benefit of
being constructed to remove contamination from the tSZ effect,
provided that the
tSZ spectral energy distribution is perfectly known across the frequency
channels. The {\tt 2D-ILC} CMB map has been produced by taking the
\Planck\ 2015 data and implementing the ``constrained ILC'' method developed in
\citet{2011MNRAS.410.2481R}. This component-separation approach was specifically
designed to cancel out in the CMB map any residual of the tSZ effect towards
galaxy clusters by using spectral filtering, as we now describe.

For a given frequency band $i$, the \Planck\ observation map $x_i$ can be
modelled as the combination of different emission components:
\begin{eqnarray}
  x_i(\vec{\hat{r}})\, =\, a_i\, s_{\rm CMB} (\vec{\hat{r}})\,
  +\, b_i\, s_{\rm tSZ} (\vec{\hat{r}})\, +\,
  n_i(\vec{\hat{r}}),
\end{eqnarray}
where $s_{\rm CMB} (\vec{\hat{r}})$ is the CMB temperature anisotropy
at pixel $\vec{\hat{r}}$, $s_{\rm tSZ} (\vec{\hat{r}})$ is the
tSZ fluctuation in the same direction, and $n_i(p)$ is a ``nuisance'' term
including instrumental noise and Galactic foregrounds at frequency $i$. The CMB
fluctuations scale with frequency through a known emission law parameterized by
the vector $\vec{a}$, with nine components, accounting for the nine
\Planck\ frequency bands. The emission law of the tSZ fluctuations is also known
and can be parameterized by the scaling vector $\vec{b}$ in the \Planck\
frequency bands. The kSZ signal is implicitly included in the CMB fluctuations,
since CMB anisotropies and kSZ fluctuations share the same spectral
signature.

Similar to the standard {\tt NILC} method \citep{2012MNRAS.419.1163B,
2013MNRAS.430..370R}, the \twodilc\ approach makes a minimum-variance-weighted
linear combination of the \Planck\ frequency maps.
Specifically ${\hat{s}_{\rm CMB} (\vec{\hat{r}}) = \vec{w}^{\textsf T}
\vec{x}(\vec{\hat{r}}) = \sum_{i=1}^{9} w_i
x_i(\vec{\hat{r}})}$, under the condition that the scalar product of the
weight vector $\vec{w}$ and the CMB scaling vector $\vec{a}$ is equal to
unity, i.e., ${\sum_{i=1}^{9} w_i a_i = 1}$, which guarantees the
conservation of CMB anisotropies in the filtering. However, \twodilc\
\citep{2011MNRAS.410.2481R} generalizes the standard {\tt NILC} method
by offering an additional constraint for the ILC weights to be orthogonal to
the tSZ emission law $\vec{b}$, while guaranteeing the conservation of the CMB
component. The \twodilc\ CMB estimate is thus given by
\begin{eqnarray}\label{eq:ILC}
  \hat{s}_{\rm CMB} (\vec{\hat{r}})\, =\, \vec{w}^{\textsf T}
  \vec{x}(\vec{\hat{r}}),
\end{eqnarray}
such that the variance of Eq.~\eqref{eq:ILC} is minimized, with
\begin{eqnarray}
  \vec{w}^{\tens T} \vec{a} = 1, & \label{cmbcon} \\
  \vec{w}^{\tens T} \vec{b} = 0. & \label{tszcon}
\end{eqnarray}
Benefiting from the knowledge of the CMB and tSZ spectral signatures, the
weights of the \twodilc\ are constructed in order to simultaneously yield unit
response to the CMB emission law $\vec{a}$ (Eq.~\eqref{cmbcon}) and zero response
to the tSZ emission law $\vec{b}$ (Eq.~\eqref{tszcon}). The residual
contamination from Galactic foregrounds and instrumental noise is controlled
through the condition (Eq.~\eqref{eq:ILC}). The exact expression
for the \twodilc\ weights was derived in \citet{2011MNRAS.410.2481R} by
solving the minimization problem (Eqs.~\eqref{eq:ILC}, \eqref{cmbcon}, and
\eqref{tszcon}):
\begin{align}\label{eq:2D-ILC}
  \hat{s}_{\rm CMB}(\vec{\hat{r}}) &=
  { \left( \vec{b}^{\textsf T} \tens{C}_{\vec{x}}^{-1} \vec{b} \right)
  \vec{a}^{\textsf T} {\tens{C}_{\vec{x}}}^{-1} - \left( \vec{a}^{\textsf T}
  \tens{C}_{\vec{x}}^{-1} \vec{b} \right) \vec{b}^{\textsf T}
  {\tens{C}_{\vec{x}}}^{-1}
  \over
  \left(\vec{a}^{\textsf T} \tens{C}_{\vec{x}}^{-1} \vec{a}
  \right)\left(\vec{b}^{\textsf T} \tens{C}_{\vec{x}}^{-1} \vec{b} \right) -
  \left( \vec{a}^{\textsf T} \tens{C}_{\vec{x}}^{-1} \vec{b} \right)^2 }\,
  \vec{x}(\vec{\hat{r}}),
\end{align}
where $\tens{C}_{\vec{x}}^{i j} = \left<\,x_{i}\, x_{j}\,\right>$
are the coefficients of the frequency-frequency covariance matrix of the
\Planck\ channel maps; in practice we compute this locally in each pixel $p$ as
\begin{eqnarray}\label{eq:cov}
  \tens{C}_{\vec{x}}^{i j}(p) = \sum_{p'\in \mathcal{D}(p)}
  x_i(p')x_j(p').
\end{eqnarray}
Here the pixel domain $\mathcal{D}(p)$ (referred to as ``super pixels'') around
the pixel $p$ is determined by using the following procedure: the product of
frequency maps $x_{i}$ and $x_{j}$ is convolved with a Gaussian kernel in pixel
space in order to avoid sharp edges at the boundaries of super pixels that
would create spurious
power \citep[][]{2012MNRAS.419.1163B,2013MNRAS.430..370R}.

Before applying the \twodilc\ filter (Eq.~\eqref{eq:2D-ILC}) to the
\Planck\ 2015 data, we first pre-process the data by performing point-source
``in-painting'' and wavelet decomposition, in order to optimize the foreground
cleaning. In each \Planck\ channel map we mask the point-sources detected
at ${\rm S/N}>5$ in the Second Planck Catalogue of Compact Sources PCCS2
\citep{planck2014-a35}. The masked pixels are then filled in by interpolation
with neighbouring pixels through a minimum curvature spline surface in-painting
technique, as implemented in \citet{2015MNRAS.451.4311R}. This pre-processing
of the point-source regions will guarantee reduction of the contamination from
compact foregrounds in the kSZ measurement.

The in-painted \Planck\ maps are then decomposed into a particular family
of spherical wavelets called ``needlets'' \citep[see,
e.g.,][]{Narcowich2006,Guilloux2007}. The needlet transform of the \Planck\
maps is performed as follows. The spherical harmonic coefficients ${a_{i,\,\ell
m}}$ of the \Planck\ channel maps $x_i$ are bandpass filtered in multipole
space in order to isolate the different ranges of angular scales in the data.
The \twodilc\ weights (Eq.~\eqref{eq:2D-ILC}) are then computed in pixel space
from the inverse spherical harmonic transform of the bandpass-filtered
${a_{i,\,\ell m}}$ coefficients. The frequency-frequency covariance matrix in
Eq.~(\ref{eq:cov}) is actually computed on the bandpass-filtered maps. In this
way, component separation is performed for each needlet scale (i.e., range of
multipoles) independently. Due to their localization properties, the needlets
allow for a filtering in both pixel space and multipole space,
therefore adapting the component-separation procedure to the local conditions
of contamination in both spaces
\citep[see][]{2009A&A.493.835D,2011MNRAS.418..467R,
2012MNRAS.419.1163B,2013MNRAS.430..370R}.

The top left panel of Fig.~\ref{fig:stacking} shows the result of stacking
$3^\circ \times 3^\circ$ patches of the {\tt NILC} \Planck\ CMB map in the
direction of known galaxy clusters, while the top right panel shows the result
of stacking the {\tt 2D-ILC} \Planck\ CMB map in the direction of the same
set of galaxy clusters.\footnote{The \Planck\ PSZ1 catalogue of galaxy
clusters from the 2013 \Planck\ data release has been used to determine the
position of known SZ clusters.} The \Planck\ SZ sample provides a very stringent
test, because these are the places on the sky where \Planck\ detected a
significant $y$ signature.  We see that stacking of the {\tt NILC} CMB map
shows a significant tSZ residual effect in the
direction of galaxy clusters. Conversely, the stacking of the {\tt 2D-ILC}
\Planck\ CMB map (right panel of Fig.~\ref{fig:stacking}) appears to show
substantially reduced tSZ residuals, due to the \twodilc\ filtering.  In the
bottom panels of Fig.~\ref{fig:stacking} we show the results of the stacking
procedure for the specific cluster catalogue that we will be using for the
main analysis in this paper (see next section for details).

The profiles
of the stacked patches are plotted in Fig.~\ref{fig:profiles}. The excess of
power due to tSZ residuals in the {\tt NILC} CMB map would clearly lead to a
significant bias in any attempt to detect the kSZ signal at the positions of
the galaxy clusters. As a baseline reference, the flux profile of the
\Planck\ 217-GHz map, stacked in the directions of these galaxy
clusters, is also plotted in Fig.~\ref{fig:profiles} (green squares). The tSZ
signal should in principle vanish in observations at 217\,GHz, since that is
effectively the null frequency for the tSZ signature;
in practice it is non-zero in the
\Planck\ 217-GHz map because of the broad spectral bandpass. In fact there
is an offset of about $20\,\mu$K in the flux profile of the stacked 217-GHz
map at the position of PSZ clusters. There is also still a residual offset in the {\tt 2D-ILC} CMB map; however,
it is smaller by a factor of about 2 than the tSZ signal in the baseline
\Planck\ 217-GHz map (see top panel of Fig.~\ref{fig:profiles}), and dramatically
better than for the \nilc\ map. This suggests that the method employed for the
\twodilc\ map was successful in removing the tSZ signal.

The residual flux of the stacked \twodilc\ CMB map in the direction of galaxy
clusters can be interpreted as the result of possibly imperfect assumptions in
the \twodilc\ filter and the exact tSZ spectral shape across the \Planck\
frequency bands. There may be several reasons behind incomplete knowledge
of the tSZ spectrum: detector bandpass mismatch; calibration
uncertainties; and also relativistic tSZ corrections. In addition, even if the
kSZ flux is expected to vanish on average when stacking inward- and
outward-moving clusters in a homogeneous universe, there is still a
potential selection bias
(since we use a selected subset of clusters for stacking) that may
result in a non-zero average kSZ residual in the offset of the {\tt 2D-ILC}
map.  Although it is not easy to estimate the size of all these effects, we are
confident that they cannot be too large because
the residual offset in the {\tt 2D-ILC} map is negligible
compared to the tSZ residuals in \Planck\ CMB maps, and smaller by a factor
of 2 with respect to the baseline \Planck\ 217-GHz map.

Regarding residual Galactic foreground contamination, we checked that the
angular power spectrum of the {\tt 2D-ILC} CMB map on the
60\,\% of the sky that is
unmasked is consistent with the angular power spectrum of the \Planck\ {\tt
SMICA\,} CMB maps. There is therefore no obvious excess of power due to
Galactic emission. We also checked the amount of residual dust contamination of the kSZ signal on small angular scales in the direction of the galaxy clusters, where dusty star-forming galaxies are present \citep{planck2016-XLIII}. Considering the \Planck\ 857-GHz map as a dust template, we scaled it across the \Planck\ frequency bands using a modified blackbody spectrum with best-fit values from \citet{planck2016-XLIII}, i.e., $\beta=1.5$ and $T=24.2$\,K. This provides dust maps at each frequency band. We then applied the ILC weights that go into the {\tt 2D-ILC} CMB+kSZ map (Eq.~(\ref{eq:2D-ILC})) to the thermal dust maps. This provides an estimate of the map of the residual dust contamination in the {\tt 2D-ILC} map. We then stacked the residual dust map in the direction of the galaxy clusters from either the PSZ or the MCXC catalogue, and computed the profile of the stacked patch as in Fig.~\ref{fig:profiles}. We found that the residual flux from the dust stacked in the direction of the galaxy clusters is compatible with zero.

Residual cosmic infrared background (hereafter CIB) and instrumental noise in the CMB maps will add some scatter to
the measured kSZ signals in the directions of galaxy clusters, but should not
lead to any bias in the stacked profile. However, any additional source of
extra noise will lead to bias in the variance of the stacked profile. Since CIB
and noise are not spatially localized on the sky (unlike kSZ and tSZ signals)
this bias can be estimated using off-cluster positions, e.g., for the
matched-filtering analysis performed in Sect.~\ref{sec:statistics}.

In order to quantify the amount of residual noise in the {\tt 2D-ILC} CMB
map, we apply the \twodilc\ weights (calculated from the \Planck\
full-survey maps) to the first and second halves of each stable pointing period
(also called ``rings''). In the half-difference of the resulting ``first'' and
``second'' {\tt 2D-ILC} maps, the sky emission cancels out, therefore leaving
an estimate of the noise contamination in the {\tt 2D-ILC} CMB maps,
constructed from the full-survey data set.

The {\tt 2D-ILC} CMB map shows approximately 10\,\% more noise than the {\tt
NILC} CMB map; this arises from the additional
constraint imposed in the {\tt 2D-ILC} of cancelling out the tSZ emission. At
the cost of having a slightly higher noise level, the {\tt 2D-ILC} CMB map
benefits from the absence of bias due to tSZ in the directions of galaxy
clusters. For this reason, the {\tt 2D-ILC} CMB map is particularly well
suited for the extraction of the kSZ signal in the direction of galaxy
clusters and we shall focus on it for the main results of this paper.

\subsection{The MCXC X-ray catalogue}
\label{sec:MCXC}

To trace the underlying baryon distribution, we use the Meta Catalogue of X-ray
detected Clusters of galaxies (MCXC), which is an all-sky
compilation of $1743$ all-sky
ROSAT survey-based samples (BCS, \citealt{eb98,eb00}; CIZA, \citealt{eb10,koc};
MACS, \citealt{eb07}; NEP, \citealt{Henry06}; NORAS, \citealt{boh00}; REFLEX,
\citealt{boh04}; SCP, \citealt{cru}) along with a few other catalogues (160SD,
\citealt{mul}; 400SD, \citealt{bure}; EMSS, \citealt{gio,Henry04}; SHARC,
\citealt{rom,burk}; WARPS, \citealt{per,hor}). We show stacks and profiles
for this catalogue on the {\it Planck} map in Figs.~\ref{fig:stacking} and \ref{fig:profiles}.
While selecting sources from this catalogue, we use the luminosity within
$R_{500}$ (the radius of the cluster within which the density is $500$ times
the cosmic critical density), $L_{500}$, and restrict the samples to have
$1.5 \times 10^{33}\,{\rm W} < L_{500}< 3.7 \times 10^{38}\,{\rm W}$ within the
band 0.1--2.4\,keV \citep[see][]{pif}. As well as $L_{500}$, for each
cluster the catalogue gives $M_{500}$, the mass enclosed within $R_{500}$ at
redshift $z$, i.e., $M_{500}=(4\pi/3)500 \rho_{\rm crit}(z)R^{3}_{500}$,
estimated using the empirical relation $L_{500} \propto M_{500}^{1.64}$
in \citet{Arnaud10}.  Further details of catalogue homogenization and
calibration are described in \cite{pif} and \cite{planck2013-XIII}.

For each cluster in the MCXC catalogue, the properties we use in the rest of
this paper are the sky position (Galactic
coordinates $l,\,b$), the redshift $z$, and the mass $M_{500}$. In
Sect.~\ref{sec:velocity} we will use $M_{500}$ and $z$ to estimate the optical
depth for each cluster.

Since in the CMB map, the Galactic plane region is highly contaminated by
foreground emission, we use the \Planck\ Galactic and point-source
mask to remove 40\,\% of the
sky area. The number of MCXC sources outside the sky mask is
$N_{\rm c}=1526$ (which we use throughout the paper) and their spatial and
redshift distributions are shown in Fig.~\ref{fig:cluster}.
The full-sky distribution is presented
in the left panel of Fig.~\ref{fig:cluster}, and one can see that the
distribution of MCXC clusters is roughly uniform outside the Galactic mask. The
redshift of MCXC clusters peaks at $z=0.09$, with a long tail towards higher
redshift, $z\ga0.4$.

\section{Methodology and statistical tests}
\label{sec:methodology}

\begin{figure}
\centering
\includegraphics[width=9.cm]{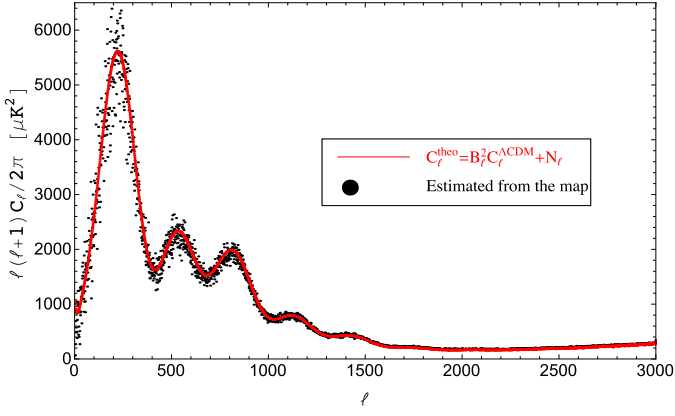}
\caption[fig:cl]{Measured (black dots) and predicted (red line) power spectra
from the \Planck\ {\tt 2D-ILC} map. The predicted spectrum is based on the
best-fitting $\Lambda$CDM model convolved with the squared beam $B^2_{\ell}$,
with the noise added. These are estimated using the pseudo-$C_{\ell}$
estimator described in \citet{Hivon02}.}
\label{fig:cl}
\end{figure}

\begin{figure}
\centering
\includegraphics[width=9.cm]{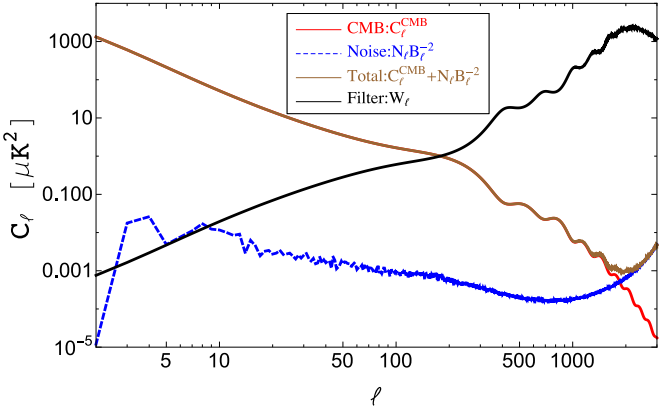}
\caption[fig:filter]{Optimal matched filter (black line) for point-source
detection in the \Planck\ {\tt 2D-ILC} map (Eq.~(\ref{eq:wlfunc})). For
comparison, the power spectra of the CMB signal (red line) and noise map
(blue dashed line) are
shown, along with their sum (brown line).}
\label{fig:filter}
\end{figure}

\begin{figure}
\centering
\includegraphics[width=9.cm]{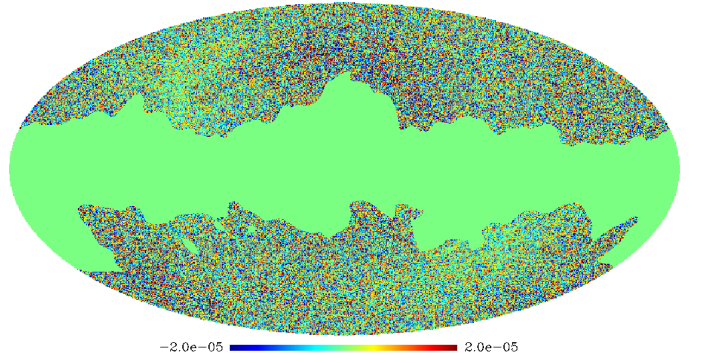}
\caption[fig:convolve]{Filtered (with Eq.~(\ref{eq:wlfunc})) and masked {\tt
2D-ILC} map in dimensionless units (i.e., $\Delta T/T$).}
\label{fig:convolve}
\end{figure}

\subsection{Matched-filter technique}
\label{sec:filter}

The foreground-cleaned CMB maps (\sevem, \smica, \nilc, \commander, and {\tt
2D-ILC}) contain mainly the primary CMB and kSZ signals, so in order to
optimally characterize the kSZ signal, we need to use a spatial filter to
convolve the maps in order to downweight the CMB signal. Here we use the
matched-filter technique \citep[e.g.,][]{Tegmark98,Ma13}, which is an easily
implemented approach for suppressing the primary CMB and instrumental noise.

Most of \Planck's SZ-clusters are unresolved, so we treat them as point
sources on the sky.  In this limit, if cluster $i$ has flux $S_i$ at sky
position $\hat{r}_i$, the sky temperature $\Delta T(\vec{\hat{r}})$ can be
written as
\begin{eqnarray}
  \Delta T(\vec{\hat{r}}) = c \sum_{j} S_j\,
  \delta_{\rm D}(\vec{\hat{r}},\vec{\hat{r}}_{j}) + \sum_{\ell m} a_{\ell m}
  Y_{\ell m}(\vec{\hat{r}}), \label{eq:deltaT}
\end{eqnarray}
where $\delta_{\rm D}$ is the Dirac delta function, $c$ is the conversion factor
between flux and temperature, and the
spherical harmonics characterize the true CMB
fluctuations. The sky signal, obtained from the \Planck\ telescope, is
\begin{eqnarray}
  \Delta T^{\rm obs}(\vec{\hat{r}}) & = & c \sum_j S_j \left( \sum_{\ell}
  \frac{2 \ell+1}{4\pi} \, P_{\ell}(\vec{\hat{r}}\cdot \vec{\hat{r}}_{j})
  \, B_{\ell} \right) \nonumber \\ &+& \sum_{\ell m} a^{\rm noise}_{\ell m}
  Y_{\ell m}(\vec{\hat{r}}),
  \label{eq:deltaT2}
\end{eqnarray}
where $a^{\rm noise}_{\ell m}$ is the true CMB signal convolved with the beam
plus the detector noise, i.e., $a^{\rm noise}_{\ell m}=B_{\ell}a^{\rm CMB}_{\ell
m}+n_{\ell m}$ (assuming that this is the only source of noise).
The beam function of \Planck\ foreground-cleaned maps in
$\ell$-space is close to a Gaussian with $\theta_{\rm
FWHM}=5\,$arcmin, i.e., $B_{\ell}=\exp(-\ell^{2}\sigma^{2}_{\rm b}/2)$, with
$\sigma_{\rm b}=\theta_{\rm FWHM}/\sqrt{8\ln 2}$.
Residual foregrounds in the \Planck\ CMB maps and in the {\tt 2D-ILC} CMB
map have been minimized in the component-separation algorithms, as demonstrated
in \citet{planck2014-a11} for the public \Planck\ CMB maps and in
\citet{planck2013-XIII} for the {\tt 2D-ILC} CMB map.
Figure~\ref{fig:cl} compares the angular power spectrum, $C_{\ell}$, directly
estimated from the map by using the pseudo-$C_{\ell}$
estimator \citep{Hivon02}, and the spectrum predicted by using the best-fit
$\Lambda$CDM model and noise template. One can see that the measured spectral
data scatter around the predicted spectrum, and that the two spectra are quite
consistent with each other.

In order to maximize our sensitivity to SZ clusters, we further convolve
$\Delta T^{\rm obs}(\vec{\hat{r}})$ with an optimal filter $W_{\ell}$:
\begin{eqnarray}
  \Delta \tilde{T}(\vec{\hat{r}}) & = & c \sum_{j} S_{j} \left(\sum_{\ell}
  \frac{2 \ell +1}{4\pi} \, P_{\ell}(\vec{\hat{r}} \cdot \vec{\hat{r}}_j)
  \, B_{\ell} \, W_{\ell} \right) \nonumber \\
  & + & \sum_{\ell m} a^{\rm noise}_{\ell m} \, W_{\ell} \, Y_{\ell
  m}(\vec{\hat{r}}),\label{eq:deltaT3}
\end{eqnarray}
where we are seeking the form of $W_{\ell}$ that will maximize cluster
signal-to-noise ratio. In the direction of each cluster, the filtered signal is
\begin{eqnarray}
  \Delta \tilde{T}_{\rm
  c}(\vec{\hat{r}}_{j})=cS_{j}\left[\sum_{\ell}\left(\frac{2 \ell +1}{4 \pi}
  \right)B_{\ell}W_{\ell} \right] \equiv (cS_{j})A, \label{eq:defA}
\end{eqnarray}
and we want to vary $W_{\ell}$ to minimize the ratio
\begin{equation}
\sigma^{2} = \textrm{Var} \left(\frac{\Delta
\tilde{T}_{\rm noise}}{\textit{A}} \right) = \frac{\sum_{\ell}
\frac{2 \ell +1}{4\pi} \, C^{\rm noise}_{\ell} \, W^2_{\ell}}{
\left(\sum_{\ell} \frac{2 \ell +1}{4\pi} \, B_{\ell} \, W_{\ell}
\right)^2}, \label{eq:sigma2}
\end{equation}
where $C^{\rm noise}_{\ell} \equiv B^2_{\ell} C^{\rm CMB}_{\ell} + N_{\ell}$, and
we take $C^{\rm CMB}_{\ell}$ to be the $\Lambda$CDM model power spectrum. Since
$A$ in Eq.~(\ref{eq:defA}) is a constant, we minimize Eq.~(\ref{eq:sigma2}) by
adding a Lagrange multiplier to the numerator \citep[see, e.g.,][]{Ma13}, i.e.,
we minimize
\begin{eqnarray}
 \sum_{\ell} \frac{2 \ell +1}{4\pi} \, C^{\rm
noise}_{\ell} \, W^2_{\ell} - \lambda \left(\sum_{\ell} \frac{2 \ell
+1}{4\pi} \, B_{\ell} \, W_{\ell} \right)^2.
\end{eqnarray}
We then obtain
\begin{eqnarray}
W_{\ell} =
\frac{B_{\ell}}{B^2_{\ell} C^{\rm CMB}_{\ell} + N_{\ell}} =
\frac{B_{\ell}}{C^{\rm noise}_{\ell}}, \label{eq:wlfunc}
\end{eqnarray}
which we plot in Fig.~\ref{fig:filter} as a black line,
along with the primary CMB $C_{\ell}$, the noise map, and their sum. In the filtering process, we use the normalized
filter $W^{\rm nor}_{\ell}=W_{\ell}/\overline{W}$, where $\overline{W}=\sum^{\ell_{\rm max}}_{\ell=1}W_{\ell}/\ell_{\rm max}$.
One can see that the filter function $W_{\ell}$ gives lower weight in the
primary CMB domain while giving more weight in the cluster regime,
$\ell\ga2000$. We then convolve the five \Planck\ foreground-cleaned maps with
this $W_{\ell}$ filter, noting that the noise power spectrum $N_{\ell}$ in
Eq.~(\ref{eq:wlfunc}) of each foreground-cleaned map is estimated by using its
corresponding noise map. After we perform this step, the primary CMB features
are highly suppressed (Fig.~\ref{fig:convolve}), and the whole sky looks
essentially like a noisy map, although it still contains the kSZ information of
course.

\subsection{Statistical method and tests of robustness}
\label{sec:statistics}

\begin{table}
\begingroup
\newdimen\tblskip \tblskip=5pt
\caption{Statistics of the values of $(\Delta T/T)\times 10^{5}$ at the true
1526 cluster positions and for 1526 randomly-selected positions.}
\label{tab:deltaTT-stats}
\nointerlineskip
\vskip -3mm
\setbox\tablebox=\vbox{
   \newdimen\digitwidth
   \setbox0=\hbox{\rm 0}
   \digitwidth=\wd0
   \catcode`*=\active
   \def*{\kern\digitwidth}
   \newdimen\signwidth
   \setbox0=\hbox{+}
   \signwidth=\wd0
   \catcode`!=\active
   \def!{\kern\signwidth}
\halign{\hbox to 1.0in{#\leaderfil}\hfil\tabskip 1em&
 \hfil#\hfil\tabskip 1em& \hfil#\hfil\tabskip 0pt\cr
\noalign{\doubleline}
\noalign{\vskip -1pt}
\omit & True positions & Random positions\cr
\noalign{\vskip 3pt\hrule\vskip 5pt}
Mean&     $-0.015$& $-0.021$\cr
Variance& $!1.38*$& $!1.23*$\cr
Skewness& $!0.37*$& $!0.09*$\cr
Kurtosis& $!4.44*$& $!3.29*$\cr
\noalign{\vskip 4pt\hrule\vskip 5pt}}}
\endPlancktable
\endgroup
\end{table}

\begin{figure}
\centering
\includegraphics[width=9.5cm]{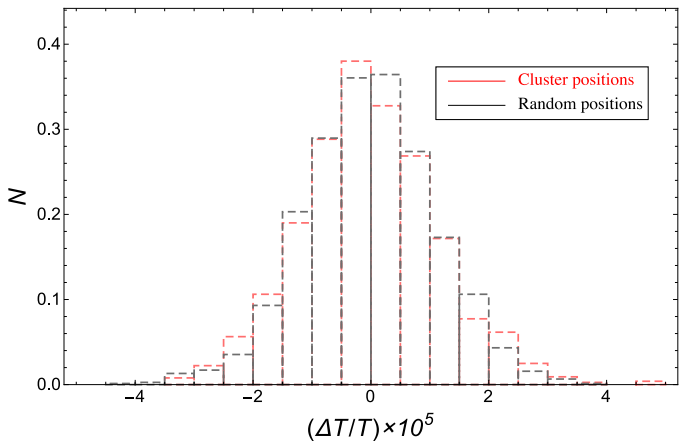}
\caption[fig:deltaTT]{The histograms of $1526$ $\Delta T/T$ values of {\tt 2D-ILC} map at
the cluster catalogue positions (red bars),
and randomly selected positions (black bars). The statistics of the true cluster positions and random positions can be found in Table~\ref{tab:deltaTT-stats}.}
\label{fig:deltaTT}
\end{figure}

\begin{table}
\begingroup
\newdimen\tblskip \tblskip=5pt
\caption{{\it Rms} values for the true sky positions of $1526$ MCXC catalogue
clusters ($\sigma_{\rm MCXC}$), along with the mean ($\sigma_{\rm ran}$) and
scatter ($\sigma(\sigma_{\rm ran})$) of the values of the \rms\ for 5000 random
catalogues, where each catalogue consists of $1526$ random positions on the
sky.} \label{tab:ran-sky}
\nointerlineskip
\vskip -3mm
\setbox\tablebox=\vbox{
   \newdimen\digitwidth
   \setbox0=\hbox{\rm 0}
   \digitwidth=\wd0
   \catcode`*=\active
   \def*{\kern\digitwidth}
   \newdimen\signwidth
   \setbox0=\hbox{+}
   \signwidth=\wd0
   \catcode`!=\active
   \def!{\kern\signwidth}
\halign{\hbox to 1.0in{#\leaderfil}\hfil\tabskip 1em&
 \hfil#\hfil\tabskip 1em& \hfil#\hfil& \hfil#\hfil\tabskip 0pt\cr
\noalign{\doubleline}
\noalign{\vskip -1pt}
\omit\hfil Map\hfil& $\sigma_{\rm MCXC}\times 10^5$&
 $\sigma_{\rm ran}\times 10^5$& $\sigma(\sigma_{\rm ran})\times 10^5$\cr
\noalign{\vskip 3pt\hrule\vskip 5pt}
{\tt 2D-ILC}& $1.17$& $1.10$& $0.022$\cr
\smica&       $1.11$& $0.97$& $0.019$\cr
\nilc&        $1.09$& $0.97$& $0.019$\cr
\sevem&       $1.12$& $1.00$& $0.020$\cr
\commander&   $1.09$& $1.03$& $0.020$\cr
\noalign{\vskip 4pt\hrule\vskip 5pt}}}
\endPlancktable
\endgroup
\end{table}

We now proceed to estimate the kSZ temperature dispersion
and perform various tests. The filtered map contains the kSZ signal and residual
noise, and from this we plot the histogram of $1526$ $\Delta T/T$ values at the
cluster positions (see red bars in Fig.~\ref{fig:deltaTT}). We can also
randomly select the same number of pixels on the sky and plot a histogram for
that. The two histograms have almost zero mean value
(Table~\ref{tab:deltaTT-stats}), but in the real cluster positions yield a
larger variance than the random selections, i.e., the real cluster positions
give a slightly broader distribution than for the randomly
selected positions (Table~\ref{tab:deltaTT-stats}). We also show results for
the skewness and kurtosis of the two samples in Table~\ref{tab:deltaTT-stats},
and one can see that for these statistics the real cluster positions also give
larger values than for the randomly-selected positions. This suggests that
there may be additional tests that could be performed to distinguish the
real cluster kSZ signals; however, we leave that for future studies, and for
the rest of this paper we just focus on investigating whether the slight
broadening of the distribution is due to the kSZ effect.

\subsubsection{Test of thermal Sunyaev-Zeldovich effect residuals}
\label{sec:tSZ-residual}

\begin{figure*}
\centering
\includegraphics[width=17.5cm]{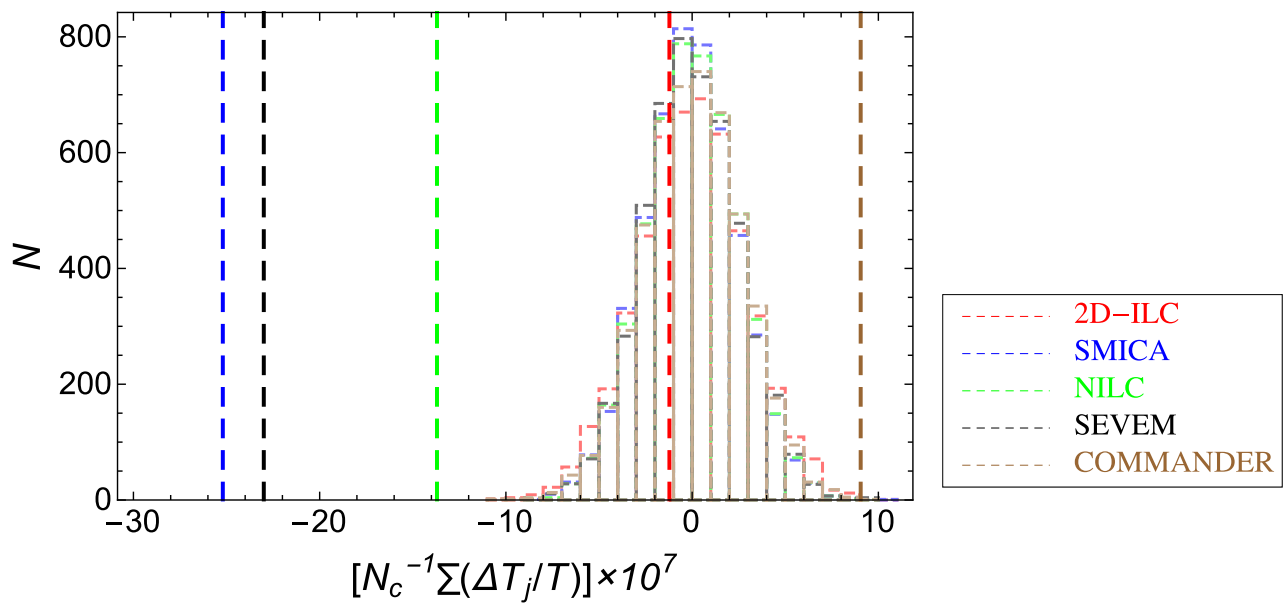}
\caption[fig:tSZ_resi]{Histogram of the $N_{\rm c}^{-1}\sum_{j}(\Delta
T_{j}/T)$ values of 5000 random catalogues on the sky (each having
$N_{\rm c}=1526$), with different colours representing different \Planck\
foreground-cleaned maps. The 68\,\%
width of the {\tt 2D-ILC}, {\tt SMICA}, {\tt NILC}, {\tt SEVEM},
and \commander\ histograms are
$2.86, 2.49, 2.48, 2.53, 2.62$ ($\times10^{-7}$), respectively. The vertical
lines represent the average $\Delta T/T$ values at true MCXC cluster positions
for each map. One can see that only for the {\tt 2D-ILC} map is this value
within the 68\,\% range of the random catalogue
distribution, while others are quite far off. This indicates that, except for
the {\tt 2D-ILC} map, the public \Planck\ maps have residual tSZ contamination
at the cluster positions.
}
\label{fig:tSZ-resi}
\end{figure*}

\begin{figure}
\centering
\includegraphics[width=9.cm]{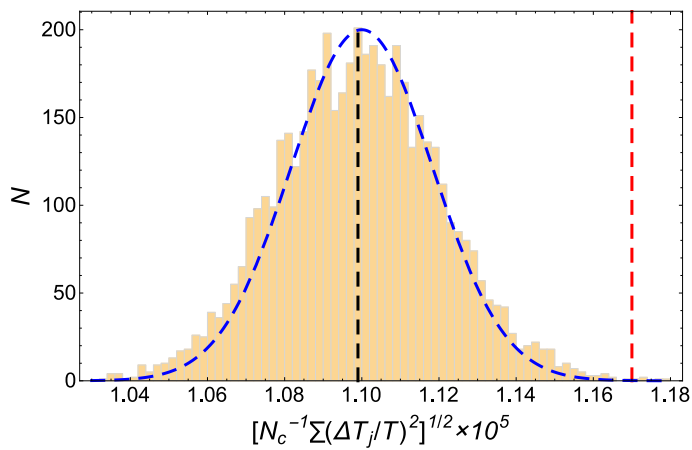}
\caption[fig:y2dilc-noise]{
Distribution of the \rms\ for 5000 mock catalogues
(yellow histogram); here each catalogue
consists of $N_{\rm c}$ randomly chosen positions on the filtered {\tt
2D-ILC} map. The mean and \rms\ of the 5000 random catalogues are $1.10
\times 10^{-5}$ (shown as the black
dashed vertical line) and $2.15 \times 10^{-7}$, respectively. For the $N_{\rm c}$ true MCXC
cluster positions, the \rms\ is $1.17 \times 10^{-5}$ (red vertical dashed line).}
\label{fig:y2dilc-noise}
\end{figure}

The first test we want to perform is to check whether the measured
kSZ $(\Delta T/T)$ value at each cluster position suffers from residuals of
the tSZ effect. The mapmaking procedures of {\tt SMICA}, {\tt NILC},
{\tt SEVEM}, and \commander\ minimize
the variance of all non-CMB contribution to the map, but they are not designed
to null the tSZ component. By contrast, the {\tt 2D-ILC} map is designed to
also null the tSZ contribution, and therefore should provide a cleaner
measurement of the kSZ effect (but with a slightly higher noise level).

We first choose 5000 randomly selected catalogues from each \Planck\
foreground-cleaned map, each being a collection of $1526$ random positions on
the sky. We then calculate the average value of $(\Delta T/T)$ for each random
catalogue and plot the resulting histograms in Fig.~\ref{fig:tSZ-resi}. The
five different colours of (overlapping) histogram represent the different \Planck\ maps.
One can see that they are all centred on zero, with approximately the same
widths. Since the {\tt 2D-ILC} map has nulled the tSZ component in the map, it
does not minimize the variance of all foreground components and as a result,
its width in Fig.~\ref{fig:tSZ-resi} ($2.86 \times 10^{-7}$) is slightly larger than for
all other maps; in these units, the $1\,\sigma$ width of the histograms for {\tt SMICA}, {\tt NILC},
{\tt SEVEM} and \commander\ are $2.49\times 10^{-7},2.48\times 10^{-7},2.53\times 10^{-7}$, and $2.62\times 10^{-7}$, respectively. This
indicates that the noise level in the filtered {\tt 2D-ILC} map is slightly
higher than for the other four maps.

We then calculate the average value of $\Delta T/T$ at the true cluster
positions for the five \Planck\ foreground-cleaned maps as the vertical bars in Fig.~\ref{fig:tSZ-resi}. One can see that
only the average value of the {\tt 2D-ILC}
map lies close to zero and within the 68\,\% width of the noise histogram,
while the values of all other maps are quite far from the centre of the
noise distribution. This strongly suggests
that at each of the true cluster positions the $(\Delta T/T)$ value contains
some contribution from the tSZ effect, so that the tSZ effect contributes
extra variance to the foreground-cleaned maps.

\subsubsection{Test with random positions}
\label{sec:random-posi-test}

We now want to test whether this slight broadening of the distribution
is a statistically significant consequence of the kSZ effect.
So for the 5000 randomly selected catalogues, we calculate the scatter
of the $1526$ ($\Delta T/T$) values.
We then plot (in Fig.~\ref{fig:y2dilc-noise}) the histogram of 5000 \rms\
values of these random catalogues, and mark the \rms\ value of the true MCXC
cluster positions for reference.  One can see that the mean of the 5000
\rms\ values of the random catalogues is $1.10 \times 10^{-5}$, and that the
scatter of the 5000 \rms\ values of the random catalogue has a width around
$2.15 \times 10^{-7}$. The \rms\ value of the $1526$ true MCXC position is
$1.17 \times 10^{-5}$, larger than the mean value at more than the $3\,\sigma$
level.

In Table~\ref{tab:ran-sky}, we list the \rms\ value for the true sky positions
of the $1526$ MCXC catalogue sources ($\sigma_{\rm MCXC}$, the mean
($\sigma_{\rm ran}$), and standard deviation
($\sigma(\sigma_{\rm ran})$) for 5000 random catalogues for different
foreground-cleaned maps. One can see that although the absolute value of each
map varies somewhat, the second column ($\sigma_{\rm ran}$) is consistently
smaller than the first column ($\sigma_{\rm MCXC}$) by roughly $0.07$--$0.13
\times 10^{-5}$, which, specifically for \twodilc\ is about $3$ times the
scatter of the \rms\ among catalogues ($\sigma(\sigma_{\rm ran})$).
This consistency strongly
suggests that the kSZ effect contributes to the extra dispersion in the
convolved $\Delta T/T$ maps at the cluster positions (since the \twodilc\ map
is constructed to remove the tSZ effect). If we use the \smica, \sevem, \nilc,
and \commander\ maps, the detailed values will vary slightly due to the
different calibration schemes of the maps, but the detection remains
consistently there.  The difference between $\sigma_{\rm MCXC}$ and
$\sigma_{\rm ran}$ is slightly larger in the \smica, \nilc, and \sevem\
maps due to residual tSZ contamination, shown as the vertical bars in Fig.~\ref{fig:tSZ-resi}.

This all points towards the broadening of the $\Delta T/T$ histogram
being a consequence of the kSZ effect; hence we identify it as additional
temperature dispersion arising from the scatter in cluster velocities
detected through the kSZ effect. In Sect.~\ref{sec:velocity}, we will interpret
this effect in terms of the line-of-sight velocity dispersion of the SZ
clusters, which is an extra variance predicted in linear
perturbation theory in the standard picture of structure formation
\citep{1980lssu.book.....P}.

\subsubsection{Test with finite cluster size}
\label{sec:cluster-size}

\begin{table}
\begingroup
\newdimen\tblskip \tblskip=5pt
\caption{Same as Table~\ref{tab:ran-sky} for the {\tt 2D-ILC} map,
but changing the assumed size of the clusters in the filtering
function~(Eq.~\eqref{eq:wlfunc}).} \label{tab:size}
\nointerlineskip
\vskip -3mm
\setbox\tablebox=\vbox{
   \newdimen\digitwidth
   \setbox0=\hbox{\rm 0}
   \digitwidth=\wd0
   \catcode`*=\active
   \def*{\kern\digitwidth}
   \newdimen\signwidth
   \setbox0=\hbox{+}
   \signwidth=\wd0
   \catcode`!=\active
   \def!{\kern\signwidth}
\halign{\hbox to 1.0in{#\leaderfil}\hfil\tabskip 1em&
 \hfil#\hfil\tabskip 1em& \hfil#\hfil& \hfil#\hfil\tabskip 0pt\cr
\noalign{\doubleline}
\noalign{\vskip -1pt}
\omit& $\sigma_{\rm MCXC}\times 10^{5}$& $\sigma_{\rm ran}\times 10^{5}$&
 $\sigma(\sigma_{\rm ran})\times 10^{5}$\cr
\noalign{\vskip 3pt\hrule\vskip 5pt}
Point source & $1.17$ & $1.10$ & $0.022$\cr
$3\,$arcmin  & $1.19$ & $1.12$ & $0.022$\cr
$5\,$arcmin  & $1.26$ & $1.19$ & $0.023$\cr
$7\,$arcmin  & $1.42$ & $1.36$ & $0.026$\cr
\noalign{\vskip 4pt\hrule\vskip 5pt}}}
\endPlancktable
\endgroup
\end{table}

We now want to test how much our results depend on the assumption that SZ
clusters are point sources. A cluster on the sky appears to have a radius of
$\theta_{500}$, which is equal to $\theta_{500}=R_{500}/D_{\rm A}$, where
$R_{500}$ is the radius from the centre of the cluster at which the density
contrast is equal to $500$ and $D_{\rm A}$ is the angular diameter distance to
the cluster. The peak in the distribution of $\theta_{500}$ values for the
MCXC clusters lies at around $3\,$arcmin, so we multiply the filter function
(Eq.~\eqref{eq:wlfunc}) with an additional ``cluster beam function''
$B^{\rm c}_{\ell}=\exp(-\ell^{2}\sigma^{2}_{\rm b}/2)$, where
$\sigma_{\rm b}=\theta_{500}/\sqrt{8\ln 2}$. We pick three different values for
the cluster size, namely $\theta_{500}=3$, $5$, and $7$\,arcmin, and see how
our results change.

We list our findings in Table~\ref{tab:size}; one can see that the detailed
values for three cases are slightly different from those of the point-source
assumptions, but the changes are not dramatic. More importantly, the offsets
between $\sigma_{\rm MCXC}$ and $\sigma_{\rm ran}$ stay the same for various
assumptions of cluster size. Therefore the detection of the temperature
dispersion due to the kSZ effect does not strongly depend on the assumption
of clusters being point sources.

\subsection{Statistical results}
\label{sec:stats-results}

\subsubsection{Statistics with the uniform weight}
\label{sec:uniform}

\begin{table}
\begingroup
\newdimen\tblskip \tblskip=5pt
\caption{Statistics of the variables $\widehat{s^{2}}$ due to the kSZ effect
for different CMB maps. } \label{tab:stats-s}
\nointerlineskip
\vskip -3mm
\setbox\tablebox=\vbox{
   \newdimen\digitwidth
   \setbox0=\hbox{\rm 0}
   \digitwidth=\wd0
   \catcode`*=\active
   \def*{\kern\digitwidth}
   \newdimen\signwidth
   \setbox0=\hbox{+}
   \signwidth=\wd0
   \catcode`!=\active
   \def!{\kern\signwidth}
\halign{\hbox to 1.0in{#\leaderfil}\hfil\tabskip 1em&
 \hfil#\hfil& \hfil#\hfil& \hfil#\hfil\tabskip 0pt\cr
\noalign{\doubleline}
\noalign{\vskip -1pt}
\omit\hfil Map\hfil& $E[s^{2}]\times10^{11}$&
 $\left(V[s^{2}]\right)^{1/2}\times10^{11}$&
 S/N\cr
\noalign{\vskip 3pt\hrule\vskip 5pt}
{\tt 2D-ILC\,}& $1.64$& $0.48$& $3.4$\cr
\smica&         $3.53$& $0.37$& $9.4$\cr
\nilc&          $2.75$& $0.38$& $7.3$\cr
\sevem&         $3.19$& $0.40$& $8.1$\cr
\commander&     $1.47$& $0.42$& $3.5$\cr
\noalign{\vskip 4pt\hrule\vskip 5pt}}}
\endPlancktablewide
\endgroup
\end{table}

\begin{table}
\begingroup
\newdimen\tblskip \tblskip=5pt
\caption{Statistics of the weighted variables $\widehat{s_{\rm w}^{2}}$ for
different choices of weights in the {\tt 2D-ILC} map. We use both linear and
squared weights for each of optical depth, luminosity, mass, and
$\theta_{500}=R_{500}/D_{\rm A}$, where $D_{\rm A}$ is the angular diameter
distance of the cluster. The third column lists the frequency
$P(s^{2}_{\rm w}<0)$ of finding a value of $s^{2}_{\rm w}$ smaller than zero,
and the fourth column lists the equivalent signal-to-noise ratio
(see Appendix~\ref{sec:P-SN}).} \label{tab:stats-weights}
\nointerlineskip
\vskip -3mm
\setbox\tablebox=\vbox{
   \newdimen\digitwidth
   \setbox0=\hbox{\rm 0}
   \digitwidth=\wd0
   \catcode`*=\active
   \def*{\kern\digitwidth}
   \newdimen\signwidth
   \setbox0=\hbox{+}
   \signwidth=\wd0
   \catcode`!=\active
   \def!{\kern\signwidth}
\halign{\hbox to 1.0in{#\leaderfil}\hfil\tabskip 1.0em&
 \hfil#\hfil& \hfil#\hfil& \hfil#\hfil& \hfil#\hfil\tabskip 0pt\cr
\noalign{\doubleline}
\noalign{\vskip -2pt}
\omit\hfil Weight \hfil& $E[s_{\rm w}^{2}]$&
 $\left(V[s_{\rm w}^{2}]\right)^{1/2}$&
$P(s^{2}_{\rm w}<0)$ & S/N\cr
\omit\hfil& $\times10^{11}$& $\times10^{11}$& & \cr
\noalign{\vskip 3pt\hrule\vskip 5pt}
Uniform&            $1.64$& $0.48$&  $*0.07$\% & $3.2$\cr
$\tau$&             $1.65$& $0.50$&  $*0.11$\% & $3.1$\cr
$\tau^{2}$&         $1.62$& $0.55$&  $*0.38$\% & $2.7$\cr
$\theta_{500}$&     $3.33$& $0.64$&  $*0.02$\% & $3.5$\cr
$\theta^{2}_{500}$& $6.86$& $1.72$&  $*0.39$\% & $2.7$\cr
$L_{500}$&          $1.34$& $0.91$&  $*6.94$\% & $1.5$\cr
$L^{2}_{500}$&      $0.65$& $2.15$&  $*32.4$\% & $0.5$\cr
$M_{500}$&          $1.91$& $0.65$&  $*0.43$\% & $2.6$\cr
$M^{2}_{500}$&      $1.81$& $1.36$&  $*8.75$\% & $1.4$\cr
\noalign{\vskip 4pt\hrule\vskip 5pt}}}
\endPlancktable
\endgroup
\end{table}

\begin{figure*}
\centerline{
\includegraphics[width=3.6in]{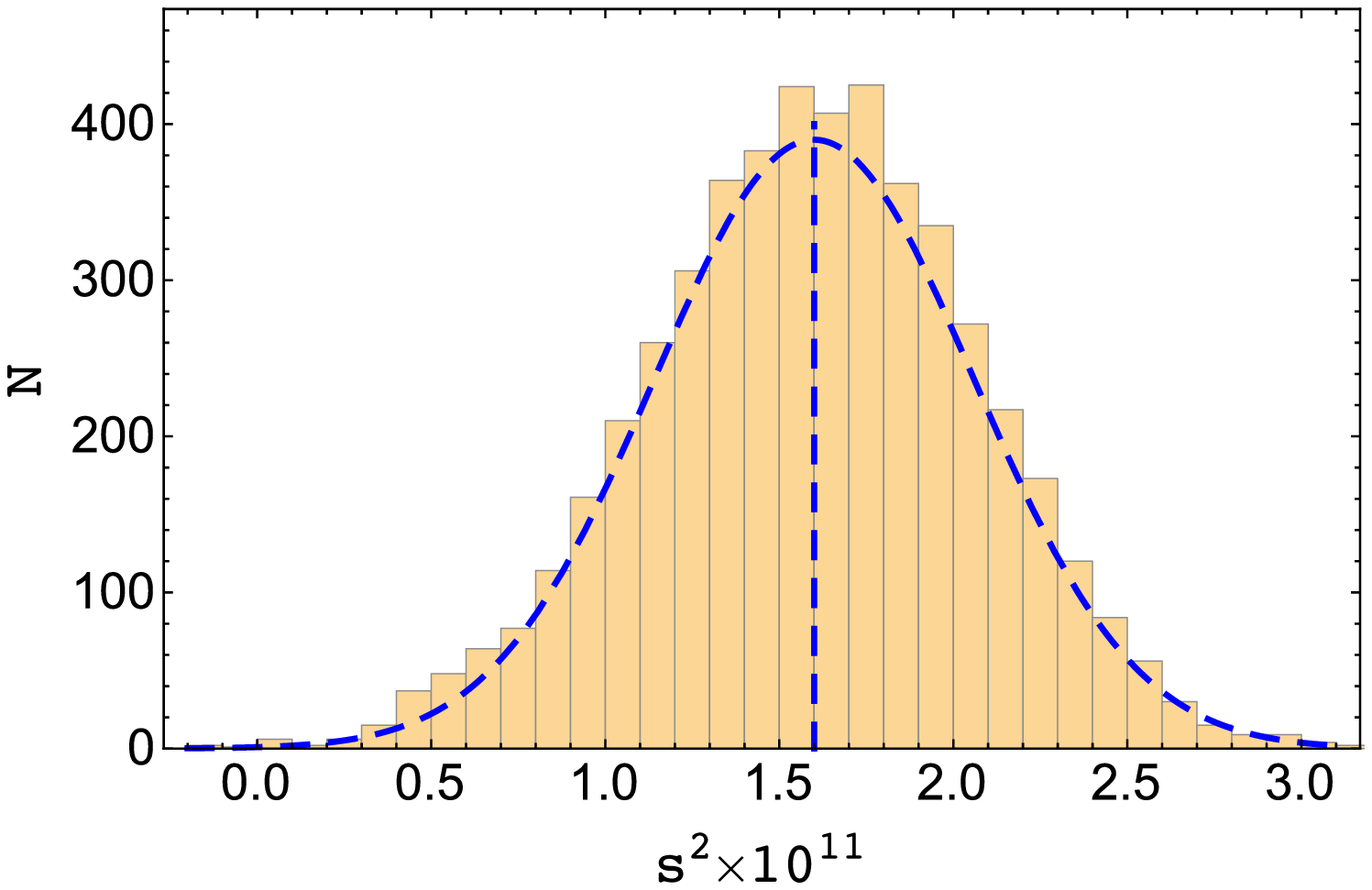}
\includegraphics[width=3.6in]{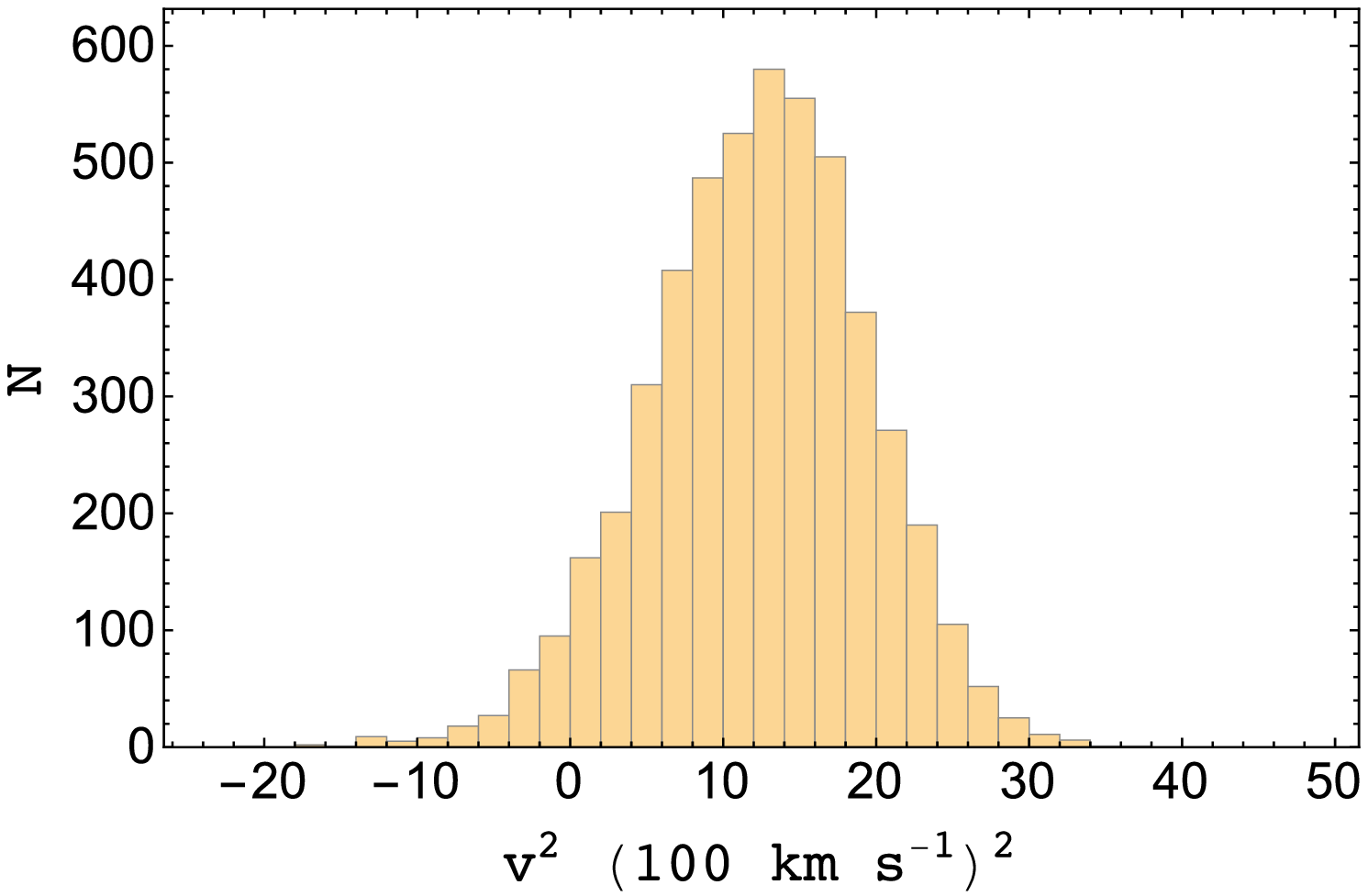}}
\caption[fig:s2-v2]{
{\it Left}: Distribution of 5000 values of $s^{2}$ with uniform weight (Eq.~\eqref{eq:def-s2}) on each position for the {\tt 2D-ILC} map.
{\it Right}: Distribution of $v^{2}$ calculated from Eq.~(\ref{eq:v-2}), after subtracting the lensing shift.  The distribution has $P(v^{2}<0)=5.3$\,\%,
corresponding to a (1-sided) detection of $1.6\,\sigma$. We have tested with 50\,000 values of $s^{2}$ and the results are consistent with those from 5000 values.}
\label{fig:s2-v2}
\end{figure*}

We now want to perform a more quantitative calculation of the significance of
detection. Since the convolved map mainly consists of the kSZ signal at the
cluster positions plus residual noise, we write the observed temperature
fluctuation at the cluster positions as
\begin{eqnarray}
\left(\frac{\Delta T}{T} \right)\equiv \delta =s + n,\label{eq:ysn}
\end{eqnarray}
where $\delta$, $s$, and $n$ represent the observed $\Delta T/T$ value, the
kSZ signal contribution, and the residual noise, respectively,
all of which are dimensionless
quantities. Now we define the estimator $\widehat{s^{2}}$ as
\begin{eqnarray}
\widehat{s^{2}}=\frac{1}{N_{\rm c}}\sum_{i} \delta_{i}^{2} -\frac{1}{N_{\rm c}} \sum_{i} \hat{n}^{2}_{i} , \label{eq:def-s2}
\end{eqnarray}
where the summation includes all of the $N_{\rm c}=1526$ cluster
positions. For the first term $\delta_{i}$, we use the $N_{\rm c}$ true cluster position as the measurement of each observed $\Delta T/T$. For the second term, we randomly select $N_{\rm c}$ pixels outside the Galactic and point-source mask that are {\it not} cluster positions. The calculation of the first term is fixed, whereas the second term depends on the $N_{\rm c}$ random positions we choose. Each randomly selected set of $N_{\rm c}$ positions corresponds to a mock catalogue, which leads to one value of $s^{2}$. We do this for 5000 such catalogues, where each mock catalogue has a different noise part ($\hat{n}_{i}$) in Eq.~(\ref{eq:def-s2}), but the same observed $\delta_{i}$. Then we plot the histogram of $s^{2}$ values for these catalogues in the left panel of Fig.~\ref{fig:s2-v2}. One can see that the $s^{2}$ distribution is close to a Gaussian distribution with mean and error being $s^{2}=(1.64 \pm 0.48) \times 10^{-11}$.

One can use a complementary method to obtain the mean and variance of $\widehat{s^{2}}$, i.e., $E[s^{2}]$ and $V[s^{2}]$. We lay out this calculation
in Appendix~\ref{sec:AppA}, where we directly derive these results:
\begin{eqnarray}
E[s^{2}] &=& \overline{\delta^{2}}-\mu_{2}(n); \nonumber \\
V[s^{2}] &=& \frac{1}{N_{\rm c}}\left[\mu_{4}(n)-\mu^{2}_{2}(n)\right].
 \label{eq:Es2-Vs2}
\end{eqnarray}
Here $\mu_{2}$ and $\mu_{4}$ are the second and fourth moments of the
corresponding random variables.

For the moments of $\delta$, we use the measurements at the $1526$ cluster positions.
For the estimate of the noise, we take all of the unmasked pixels of the
convolved sky. In order to avoid selecting the real cluster positions,
we remove all pixels inside a $10\,$arcmin aperture around each cluster. These
``holes'' at each cluster position constitute a negligible portion of the total
unmasked pixels, and our results are not sensitive to the aperture size we
choose. As a result, we have approximately $3\times 10^{7}$ unmasked pixels
to sample the noise. We then substitute the values into
Eq.~(\ref{eq:Es2-Vs2}) to obtain the expectation values
and variances.

In Table~\ref{tab:stats-s}, we list the mean and \rms\ value of
$\widehat{s^{2}}$. Comparing with the {\tt 2D-ILC} map, one can see that
the \smica, \nilc, and \sevem\ maps give larger values of $E[s^{2}]$ and
therefore apparently higher significance levels,
which we believe could be due to the fact that the residual tSZ effect in
these maps contributes to the signal. However, the \commander\
map gives a reasonable estimate of the dispersion, since it appears to be less
contaminated by tSZ residuals (see Fig.~\ref{fig:tSZ-resi}).  As discussed
in Sect.~\ref{sec:2DILC}, the mapmaking procedure of the {\tt 2D-ILC} product
enables us to null the tSZ effect so that the final map should be free of tSZ,
but with larger noise. This is the reason that we obtain a somewhat lower
significance in Table~\ref{tab:stats-s} for {\tt 2D-ILC} compared to some
of the other maps.  We will therefore mainly quote this
conservative detection in the subsequent analysis.

\subsubsection{Statistics with different weights}
\label{sec:diff-weight}

The results so far have been found using the same weights for each cluster
position. We now examine the stability of the detection using
weighted stacking. In Eq.~(\ref{eq:def-s2}) we defined stacking with uniform
weights, which can be generalized to
\begin{eqnarray}
\widehat{s}^{2}_{\rm w}= \frac{\sum_{i}\left(\delta^{2}_{i} -\hat{n}^{2}_{i} \right) w_{i}}{\sum_{i} w_{i}} , \label{eq:def-sw}
\end{eqnarray}
where $w_{i}$ is the weight function. We certainly expect ``larger'' clusters
to contribute more to the signal, but it is not obvious what cluster property
will be best to use.  In Table~\ref{tab:stats-weights}, we try different
weighting functions $w_{i}$, with the first row being the uniform weight, which
is equivalent to Eq.~(\ref{eq:def-s2}). In addition, we try as different
choices of weighting function the optical depth $\tau$ and its square
$\tau^{2}$,\footnote{The calculation of optical depth is shown in
Sect.~\ref{sec:velocity}, and Eq.~(\ref{eq:tau2}) in particular.} the angular
size $\theta_{500}$ and its square $\theta^{2}_{500}$, the luminosity $L_{500}$
and its square $L^{2}_{500}$, and the mass $M_{500}$ and its square
$M^{2}_{500}$. Since some of these may give distributions of $s^{2}_{\rm w}$
that deviate from Gaussians, we also calculate the frequencies for finding
$s^{2}_{\rm w}$ smaller than zero, $P(s^{2}_{\rm w})<0$. The smaller this
$P$-value is, the more significant is the detection.

From Table~\ref{tab:stats-weights}, we see that most weighting choices are
consistent with uniform weighting though with reduced
significance of the detection, the exceptions being the choices of
$\theta_{500}$ or $\theta_{500}^2$. For $w_{i}=\theta_{500, i}$ and
$w_{i}=\theta^{2}_{500, i}$, we have $P$-values of $0.0002$ and $0.0039$,
respectively, yielding (1-sided) significance levels of $3.5\,\sigma$ and
$2.7\,\sigma$. Their distributions deviate slightly from a Gaussian, with a
tail toward smaller values.

The increased detection of excess variance using $\theta_{500}$ weighting stems
from our choice of using a single cluster beam function for all clusters. In
Sect.~\ref{sec:cluster-size} we tested the robustness of our results to the
choice of cluster beam function, finding little dependence. Nevertheless, such
a test assumed all clusters had the same angular size, while in
reality there is a large spread in the angular sizes of the clusters.
By weighting with $\theta_{500}$ we are able to
recover some of this lost signal in a quick and simple way, which we tested
by comparing results for larger clusters versus smaller clusters.
Despite this, we find that the increased significance is mainly due to the
increased value for $E[s^2_w]$ and not a decrease in the noise. This tension may be evidence of systematic effects in the data or potentially some noise coming from residual tSZ signal in the 2D-ILC map; this should be
further investigated when better data become available.

\subsubsection{Statistics with split samples}

\begin{table}
\begingroup
\newdimen\tblskip \tblskip=5pt
\caption{Statistics of split samples. Here the median values of the samples
are $\left(M_{500}\right)_{\rm mid}=1.68\times 10^{14}\,{\rm M}_{\odot}$,
$\left(L_{500}\right)_{\rm mid}=2.3\times 10^{10}\,{\rm L}_\odot$,
and $\left(R_{500}\right)_{\rm mid}=0.79\,{\rm Mpc}$. The number of sources in
each split sample is $763$.} \label{tab:stats-split}
\nointerlineskip
\vskip -3mm
\setbox\tablebox=\vbox{
   \newdimen\digitwidth
   \setbox0=\hbox{\rm 0}
   \digitwidth=\wd0
   \catcode`*=\active
   \def*{\kern\digitwidth}
   \newdimen\signwidth
   \setbox0=\hbox{+}
   \signwidth=\wd0
   \catcode`!=\active
   \def!{\kern\signwidth}
\halign{\hbox to 1.5in{#\leaderfil}\hfil\tabskip 0.5em&
 \hfil#\hfil&  \hfil#\hfil& \hfil#\hfil& \hfil#\hfil\tabskip 0pt\cr
\noalign{\doubleline}
\noalign{\vskip -2pt}
\omit\hfil Criterion\hfil& $E[s_{\rm w}^{2}]$&
 $\left(V[s_{\rm w}^{2}]\right)^{1/2}$&
$P(s^{2}_{\rm w}<0)$ & S/N\cr
\omit\hfil&  $\times10^{11}$& $\times10^{11}$& & \cr
\noalign{\vskip 3pt\hrule\vskip 5pt}
\,\,$L_{500}<\left(L_{500}\right)_{\rm mid}$& $0.78$& $0.68$& $12.7$\%& $1.1$\cr
\,\,$L_{500}>\left(L_{500}\right)_{\rm mid}$& $2.48$& $0.68$& $0.07$\%& $3.2$\cr
    $M_{500}<\left(M_{500}\right)_{\rm mid}$& $0.83$& $0.68$& $11.4$\%& $1.2$\cr
    $M_{500}>\left(M_{500}\right)_{\rm mid}$& $2.45$& $0.68$& $0.07$\%& $3.2$\cr
  \,$R_{500}<\left(R_{500}\right)_{\rm mid}$& $0.48$& $0.68$& $23.7$\%& $0.7$\cr
  \,$R_{500}>\left(R_{500}\right)_{\rm mid}$& $2.80$& $0.68$& $0.02$\%& $3.5$\cr
\noalign{\vskip 4pt\hrule\vskip 5pt}}}
\endPlancktable
\endgroup
\end{table}

\begin{figure*}
\centering
\includegraphics[width=9.cm]{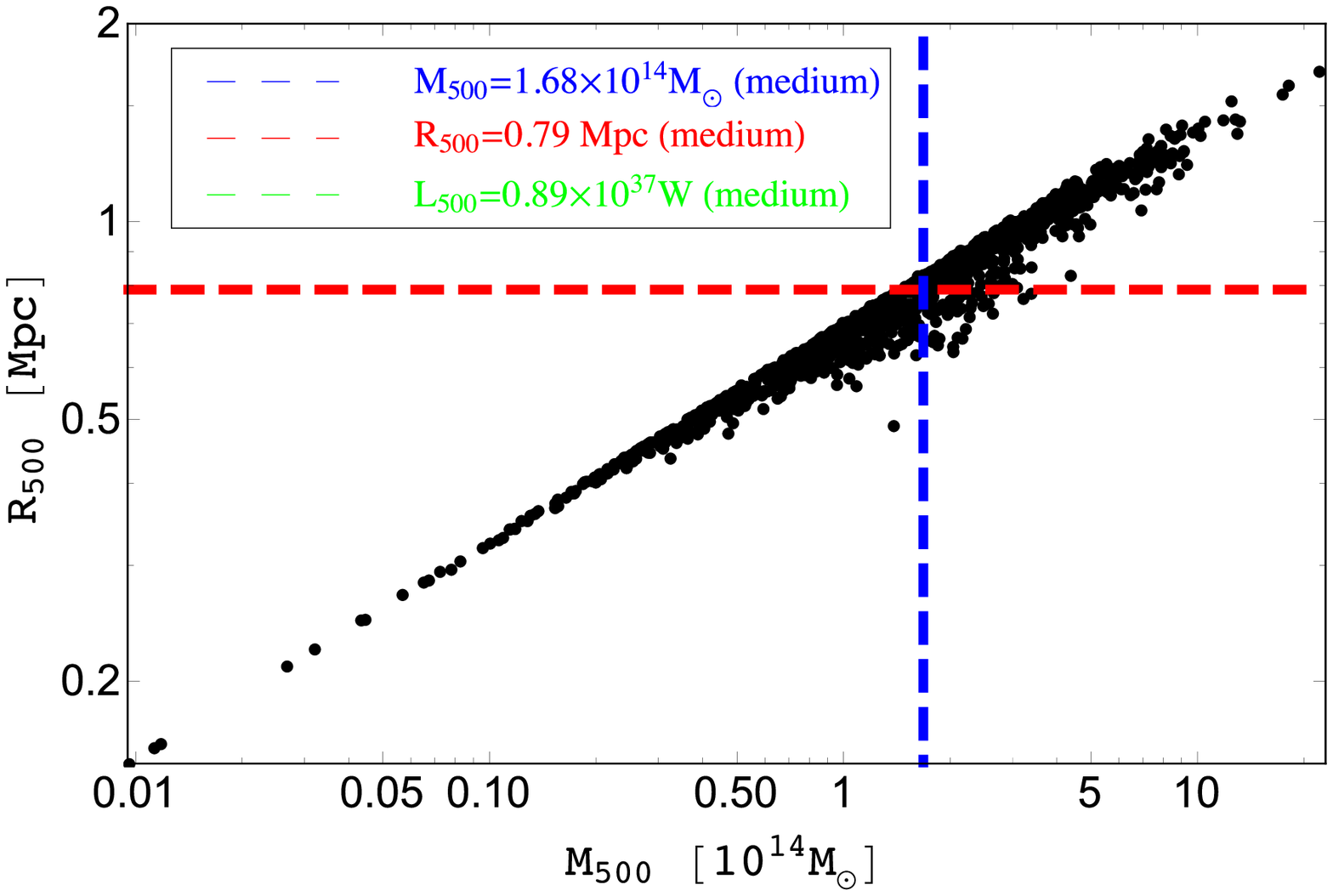}
\includegraphics[width=8.5cm]{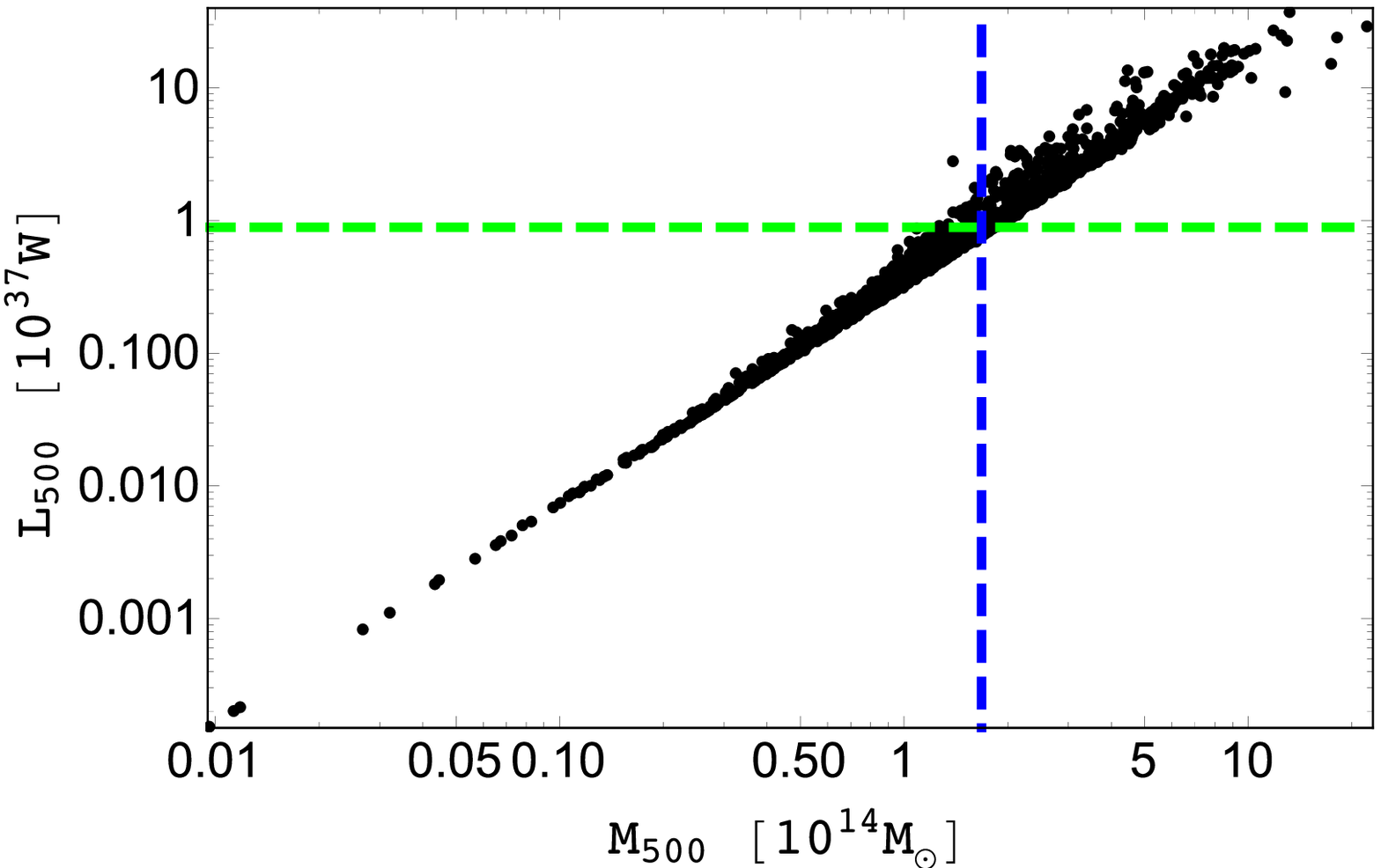}
\caption[fig:split]{{\it Left}: $R_{500}$ versus $M_{500}$ for the sample.
{\it Right}: $L_{500}$ versus $M_{500}$ for the sample. The vertical and
horizontal lines show the median values of $R_{500}$, $M_{500}$,
and $L_{500}$.}
\label{fig:split}
\end{figure*}

We further split the samples into two groups, according to their median values
of $M_{500}$, $R_{500}$ and $L_{500}$. The distributions of these values are
shown in Fig.~\ref{fig:split}. One can see that our samples span several orders
of magnitude in mass, radius, and luminosity, but that the median values of the
three quantities quite consistently split the samples into two.
We re-calculate the $\widehat{s^{2}}$ statistics as defined in
Eq.~(\ref{eq:def-s2}) for the two sub-samples and list our results in
Table~\ref{tab:stats-split}.

We see that the groups with lower values of $L_{500}$, $M_{500}$, and
$R_{500}$ give detections of the kSZ temperature dispersion at $1.1\,\sigma$,
$1.2\,\sigma$, and $0.7\,\sigma$, whereas the higher value subsets give higher
S/N detections. It is clear that the signal we see is dominated by the
larger, more massive, and more luminous subset of clusters, as one might
expect. In principle one could use this information (and that of the previous
subsection) to devise an estimator that takes into account the variability of
cluster properties in order to further maximize the kSZ signal; however, a
cursory exploration found that, for our current data, improvements will not
be dramatic.  We therefore leave further investigations for future work.

\subsection{Effect of lensing}
\label{sec:lensing}

An additional effect that could cause a correlation between our tSZ-free CMB
maps and clusters comes from gravitational lensing
\citep[as discussed in][]{Ferraro16}.  It is therefore important to determine
what fraction of our putative kSZ signal might come from lensing, and we
estimate this in the following way.
The MCMX CMB temperature variance is set by the integrated local CMB power
spectrum at these positions. Lensing by the clusters magnifies the CMB, locally
shifting scales with respect to the global average, potentially introducing a
lensing signal in the variance shift. If we were to compare the CMB variance
between cluster lensed and unlensed skies unlimited by resolution, no
difference would be seen; lensing is merely a remapping of points on the sky,
and hence does not affect one-point measures, with local shifts in scales in
the power spectrum being compensated by subtle changes in amplitude, keeping
the variance the same. However, as noted by \cite{Ferraro16}, the presence of
finite beams and of the filtering breaks this invariance and the relevance of
the lensing effects needs to be assessed. Crudely, the cluster-lensing signal
$\boldsymbol{\alpha} \cdot \nabla T$ (with $\boldsymbol{\alpha}$ the deflection
due to the cluster) can be as large as $5\,\muK$~\citep{Lewis:2006fu},
hence potentially contributing $\simeq25\,\mu{\rm K}^2$ to the observed
variance $\hat{s^2} = (121 \pm 35)\muK^{2}$, i.e., around the $1\,\sigma$ level.

\begin{figure}
\centering
\includegraphics[width=9.cm]{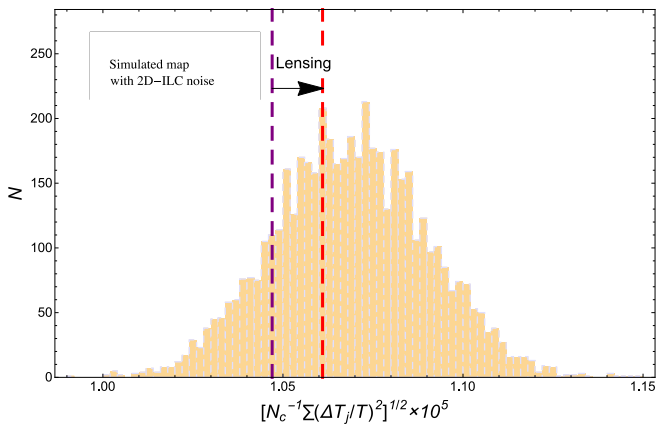}
\caption[fig:lensing]{Histogram of the $[N_{\rm c}^{-1} \sum_{j}(\Delta T_{j}/T)^{2}]^{1/2}$ values of 5000 random catalogues (each having $N_{\rm c}=1526$) on simulated lensed skies with added {\tt 2D-ILC} noise. The mean and \rms\ for the 5000 mock catalogues are $1.07 \times 10^{-5}$ and $2.13 \times 10^{-7}$, respectively. The vertical dashed lines indicate the values at the MCXC cluster positions before and after the cluster-lensing effect is added; the values are $1.047 \times 10^{-5}$ before (purple line) and $1.061 \times 10^{-5}$ after (red line).}
\label{fig:lensing}
\end{figure}

We obtain a more precise estimate with a CMB simulation as follows.
After generating a Gaussian CMB sky with lensed CMB spectra, we lens the CMB
at each cluster position according to the deflection field predicted for a
standard halo profile \citep{Navarro96,Dodelson:2004as} of the observed redshift and
estimated mass. We use for this operation a bicubic spline
interpolation scheme, on a high-resolution grid of $0.4$ arcmin
($N_{\rm side}=8192$), using the {\tt python} lensing
tools.\footnote{Available at \url{https://github.com/carronj/LensIt}\, .}
We then add the 2D-ILC noise-map estimate to the CMB.
We can finally compare the temperature variance at the cluster positions before
and after cluster lensing. We show our results in Fig.~\ref{fig:lensing}. The
background (yellow) histogram shows the value of the dispersion of the 5000
random catalogues on the simulated map with {\tt 2D-ILC} noise added. The
purple (red) line indicates the value of dispersion of the map without (with)
cluster lensing added. One can see that with cluster lensing added the value of
dispersion is shifted to a higher value by $1.4 \times 10^{-7}$, which is
somewhat less than the \rms\ of the random catalogue
($2.1 \times 10^{-7}$). This indicates that the lensing causes a roughly
$0.7\,\sigma$ shift in the width of histogram. From Fig.~\ref{fig:lensing},
we can calculate the lensing-shifted
$\hat{s^{2}}$ as
\begin{eqnarray}
\left(\hat{s^{2}} \right)_{\rm lens} &=&\left(1.061 \times 10^{-5} \right)^{2}-\left(1.047 \times 10^{-5} \right)^{2} 
\nonumber \\
&=& 2.95 \times 10^{-12}.
\end{eqnarray} 
Therefore, using uniform weights, by subtracting the
$(\hat{s^{2}})_{\rm lens}$, the temperature dispersion $\hat{s^{2}}$ in
the {\tt 2D-ILC} map is measured to be
\begin{eqnarray}
\left(\hat{s^{2}} \right)=(1.35 \pm 0.48) \times 10^{-11},
\end{eqnarray}
which is thus detected at the $2.8\sigma$ level.

\subsection{Comparison with other kSZ studies}
\label{sec:compare}

Table~\ref{tab:history} summarizes all of the previous measurements of the kSZ
effect coming from various cross-correlation studies.  One can see that many
investigations have used the pairwise temperature-difference estimator
\citep{Handetal2012}, which was inspired by the pairwise momentum estimator
\citep{ferreiraetal99}.  However, in a different approach, recently
\citet{Hill16} and \citet{Ferraro16} cross-correlated the squared kSZ field
with the WISE galaxy projected density catalogue and obtained a roughly
$4\,\sigma$
detection. This approach has similarities with the temperature dispersion that
we probe in this paper, but is different in several ways.  Firstly, in terms
of tracers, \citet{Hill16} used the density field of the WISE-selected
galaxies, of which only the radial distribution ($W_{\rm g}(\eta)$ kernel) is
known, whereas in this work we use the MCXC X-ray catalogue in  which each
cluster's exact position and redshift are already determined.  Secondly, for
the optical depth treatment \citet{Hill16} used 46 million WISE galaxies, so
the cross-correlation with the kSZ$^2$ field contains the contribution of
diffuse gas as well as virialized gas in groups; hence they assumed that their
galaxies are tracing the velocity field with a uniform optical depth
approximation (in their notation it is the $f_{\rm gas}$ parameter). In this
paper, on the other hand, we are explicitly probing the velocity dispersion
around galaxy clusters, so our optical depth comes from each individual
cluster.  Thirdly, in \citet{Hill16} and \citet{Ferraro16} the angular scales
of the kSZ$^2$ and WISE projected density field correlation lie in the range
$\ell=400$--3000, whereas in this work, the dispersion effectively comes from
a narrower range of scales around 5\arcmin--10\arcmin.  Lastly, the
lensing contamination in our case is below $1\,\sigma$ of our signal, while in
\citet{Hill16} and \citet{Ferraro16}, the lensing is correlated with the WISE
projected density field, and so the detected signal has a larger lensing
contribution \citep[figure 1 in][]{Hill16}.

It is evident therefore that the methods discussed in this paper and described
in \citet{Hill16} are complementary.  As the mapping of the small-scale CMB
sky continues to improve, we can imagine that the assumptions made in either
approach will need to be revisited and more sophisticated methods will be
needed to probe the kSZ statistics more thoroughly, e.g., through direct
comparison with simulations.

\section{Implications for the peculiar velocity field}
\label{sec:velocity}

\begin{figure}
\centering
\includegraphics[width=9.cm]{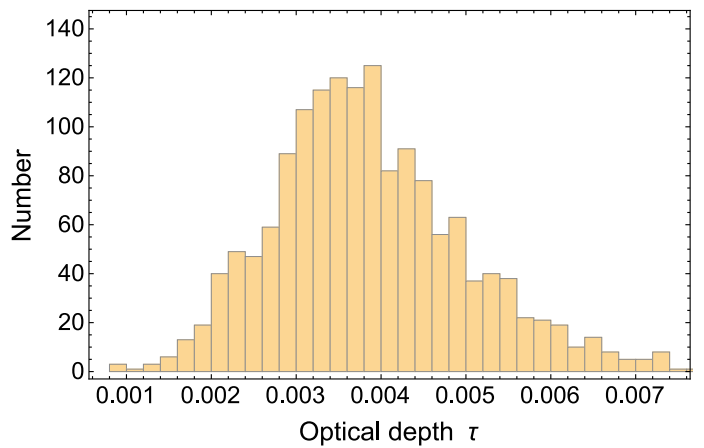}
\caption[fig:tau]{Histogram of the optical depth $\tau$ derived using
Eq.~(\ref{eq:tau2}) for $1526$ X-ray cluster positions.}
\label{fig:tau}
\end{figure}

We now want to investigate what the temperature dispersion indicates for the
variance of the peculiar velocity field. As shown in Eq.~(\ref{eq:kSZ1}), the
dimensionless temperature fluctuation is different from the dimensionless
velocity field through the line-of-sight optical depth factor $\tau$. Since the coherence
length of the velocity field is order $100\,h^{-1}\,{\rm Mpc}$
\citep{planck2015-XXXVII}, i.e., much larger than the size of a cluster,
the velocity can be taken out of the integral,
giving
\begin{eqnarray}
  \frac{\Delta T(\vec{\hat{r}})}{T} =\left(-\frac{\vec{v}\cdot
  \vec{\hat{r}}}{c} \right)\tau, \,\,\, {\rm with}\ \tau=\sigma_{\rm
  T}\int^{+\infty}_{-\infty} n_{\rm e}\der l.
\end{eqnarray}
In order to convert the kSZ signal into a line-of-sight velocity we
therefore need to obtain an estimate of $\tau$ for each
cluster. In \cite{planck2013-XIII}, the values calculated are explicitly given
as the optical depth per solid angle, obtained based on two scaling
relations from \cite{Arnaud05} and \cite{Arnaud10}. Here we adopt a slightly
different approach, which is to determine the $\tau$ value at the central
pixel of each galaxy cluster.

Many of the previous studies of the tSZ effect have used the ``universal
pressure profile'' \citep[UPP,][]{Arnaud10,planck2012-V} and isothermal
$\beta$ model \citep{Cavaliere76,Cavaliere78} to model the pressure and
electron density profiles of the
clusters \citep{Grego00,bensonetal03,Benson04,Hallman07,Halverson09,Plagge10}.
Because the UPP is just a fitting function for pressure, it is difficult to
separate out the electron density and the temperature unless we use
the isothermal assumption. In fact,  \citet{Battaglia12} demonstrated that the
UPP is not absolutely universal, and that feedback from an active galactic
nucleus can change the profile in a significant way. The functional form of the
$\beta$ model can be derived from a parameterization of density under the
assumption of isothermality of the profile \citep[e.g.,][]{Sarazin86}.
However, since isothermality is a poor assumption for many
clusters \citep{planck2012-V}, we only consider the $\beta$ model here as a
fitting function. Measurements of cluster profiles from the South Pole
Telescope (SPT) have found that the index $\beta=0.86$ provides the best fit
to the profiles of SZ clusters \citep{Plagge10}, and therefore we use this value
of $\beta$ in the following discussion.

The electron density can be written as
\begin{eqnarray}
n_{\rm e}(r)=\frac{n_{\rm e0}}{\left[1+\left(r/r_{\rm c} \right)^{2}
 \right]^{3\beta /2}},
\end{eqnarray}
where $r_{\rm c}=r_{\rm vir}/c$ is the core radius of each cluster, with $c$
being the concentration parameter. Here we adopt the formula from \cite{Duffy08} and \cite{Komatsu11} to calculate the concentration parameter given the redshift and halo mass of the cluster:
\begin{eqnarray}
c=\frac{5.72}{(1+z)^{0.71}}\left(\frac{M_{\rm vir}}{10^{14}h^{-1}{\rm M}_{\odot}} \right)^{-0.081} \label{eq:c-para}.
\end{eqnarray}
In the catalogue, $M_{500}$ and redshift $z$ are given, so one can use these two quantities to calculate the virial mass $M_{\rm vir}$ of the cluster. The calculation is contained in Appendix~\ref{sec:m500-mvir}.

The radius $r_{\rm vir}$ is calculated through
\begin{eqnarray}
M_{\rm vir}=\frac{4 \pi}{3} \left[\Delta(z)\rho_{\rm c}(z) \right]r^{3}_{\rm vir}, \label{eq:M-vir}
\end{eqnarray}
where $\rho_{\rm c}(z)$ is the critical density of the Universe at redshift
$z$, and $\Delta(z)$ depends on $\Omega_{\rm m}$ and $\Omega_{\Lambda}$
as \citep{bryan1998}
\begin{eqnarray}
\Delta(z)=18\pi^{2} + 82[\Omega(z)-1] - 39[\Omega(z)-1]^{2}, \label{eq:Delta-z}
\end{eqnarray}
with $\Omega(z)=\Omega_{\rm m}(1+z)^2\Big/\left[\Omega_{\rm m}(1+z)^{3}+\Omega_{\Lambda} \right]$.
Thus,
\begin{eqnarray}
\tau &=& (\sigma_{\rm T}\,n_{\rm e0}\,r_{\rm c})f_{1}(\beta), \nonumber \\
f_{1}(\beta) &=& \int^{+\infty}_{-\infty}\frac{\der x}{(1+x^{2})^{3 \beta /2}}=\frac{\sqrt{\pi}\,\Gamma\left(-\frac{1}{2}+\frac{3}{2}\beta \right)}{\Gamma\left(\frac{3}{2}\beta \right)}, \label{eq:tau1}
\end{eqnarray}
where $\Gamma$ is the usual gamma function. To determine $n_{\rm e0}$,
we use $4\pi \int^{r_{500}}_{0} n_{\rm e}(r) r^{2} \der r=N_{\rm e}$, where
\begin{eqnarray}
N_{\rm e}=\left(\frac{1+f_{\rm H}}{2 m_{\rm p}} \right) f_{\rm gas}M_{500}. \label{eq:Ne}
\end{eqnarray}
Here the quantity $f_{\rm H}=0.76$ is the hydrogen mass fraction,
$m_{\rm p}$ is the
proton mass, and $f_{\rm gas}=(\Omega_{\rm b}/\Omega_{\rm m})$ is the cosmic
baryon fraction, while $M_{500}$ is the cluster mass enclosed in the radius
$r_{500}$. Thus,
\begin{eqnarray}
n_{\rm e0} &=& \frac{N_{\rm e}}{4\pi r^{3}_{\rm c}f_{2}(c_{500},\beta)} \nonumber, \\
 f_{2}(c_{500},\beta) &=& \int^{c_{500}}_{0}\frac{x^{2} \der x}{(1+x^{2})^{3 \beta /2}}, \label{eq:ne0}
\end{eqnarray}
where $c_{500}=r_{500}/r_{\rm c} \simeq c_{\rm vir}/2.0$ is the concentration
parameter for $R_{500}$.

Combining Eqs.~(\ref{eq:tau1}), (\ref{eq:Ne}), and (\ref{eq:ne0}), we have
\begin{eqnarray}
\tau=\left(\frac{\sigma_{\rm T}}{4\pi r^{2}_{\rm c}} \right)\left(\frac{f_{1}(\beta)}{f_{2}(c_{500},\beta)} \right)\left(\frac{1+f_{\rm H}}{2 m_{\rm p}} \right)f_{\rm gas}M_{500}. \label{eq:tau2}
\end{eqnarray}
In Fig.~\ref{fig:tau}, we plot the histogram of the optical depth values
of the $1526$ clusters in the sample.
The mean and standard deviation are given by $\tau=(3.9 \pm 1.2) \times
10^{-3}$. This is very consistent with the quoted value $\tau =(3.75 \pm 0.89) \times 10^{-3}$ from the cross-correlation between the kSZ SPT data and the photometric data from DES survey by fitting to a template of pairwise kSZ field \citep{Soergel16}. Note that the uncertainty quoted here describes the scatter in the
mean $\tau$ values for the whole of the sample.

\begin{table}
\begingroup
\newdimen\tblskip \tblskip=5pt
\caption{Statistics of the line-of-sight velocity
dispersion $v^{2} \equiv (\vec{v}\cdot \vec{\hat{n}})^2$.} \label{tab:stats-v}
\nointerlineskip
\vskip -3mm
\setbox\tablebox=\vbox{
   \newdimen\digitwidth
   \setbox0=\hbox{\rm 0}
   \digitwidth=\wd0
   \catcode`*=\active
   \def*{\kern\digitwidth}
   \newdimen\signwidth
   \setbox0=\hbox{+}
   \signwidth=\wd0
   \catcode`!=\active
   \def!{\kern\signwidth}
\halign{\hbox to 1.0in{#\leaderfil}\hfil\tabskip 1em&
 \hfil#\hfil\tabskip 1em& \hfil#\hfil& \hfil#\hfil\tabskip 0pt\cr
\noalign{\doubleline}
\noalign{\vskip -1pt}
\omit\hfil Map\hfil& $E[v^{2}]$& $\left(V[v^{2}]\right)^{1/2}$& S/N\cr
\noalign{\vskip 3pt}
\omit& $(100\,{\rm km}\,{\rm s}^{-1})^{2}$ & $(100\,{\rm km}\,{\rm s}^{-1})^{2}$ & \cr
\noalign{\vskip 3pt\hrule\vskip 5pt}
{\tt 2D-ILC}& $12.3$& $7.1$& $1.7$\cr
\smica&         $27.0$& $5.6$& $4.8$\cr
\nilc&          $26.1$& $5.6$& $4.7$\cr
\sevem&         $23.8$& $5.9$& $4.0$\cr
\commander&     $13.5$& $6.3$& $2.1$\cr
\noalign{\vskip 4pt\hrule\vskip 5pt}}}
\endPlancktable
\endgroup
\end{table}

We convert the temperature dispersion data listed in Table~\ref{tab:stats-s} to the line-of-sight velocity dispersion measurement by using the modelled value of $\tau$, after correcting by our estimate of the lensing effect.
Our procedure is as follows. For each estimate of $\widehat{s^{2}}$, we correct
by the estimated shift caused by lensing, then calculate its $v^{2}$ value, and obtain an averaged value of $v^{2}$ via
\begin{eqnarray}
v^{2}=\frac{c^{2}}{N_{\rm c}} \sum^{N_{\rm c}}_{i=1} \frac{s^{2}_{i}}{\tau^{2}_{i}}. \label{eq:v-2}
\end{eqnarray}
We then do this for the 5000 values of $\widehat{s^{2}}$, and plot the distribution (after shifting by the lensing effect) in the right panel of Fig.~\ref{fig:s2-v2} for the {\tt 2D-ILC} map. We also present results for $v^{2}$ in Table~\ref{tab:stats-v}, where we can see
that for the conservative case, i.e., the {\tt 2D-ILC} map, the velocity
dispersion is measured to be
$v^{2}=(12.3\pm 7.1)\times
(100\,\kms)^{2}$. From the right panel of Fig.~\ref{fig:s2-v2}, we can see that
the distribution is not completely Gaussian, but has a tail towards smaller
$v^{2}$. The frequency $P(v^{2}<0)$ is 5.3\,\%, which would correspond to a
detection of the dispersion of peculiar velocity from $1526$ MCXC clusters at
the $1.7\,\sigma$ level (using Appendix~\ref{sec:P-SN}).

In studies of peculiar velocity fields, the most
relevant quantity is the linear line-of-sight velocity ($v$), or in other
words $\left\langle v^2\right\rangle^{1/2}$. We find\footnote{68\,\% CL,
although we caution that the distribution is not Gaussian}
$\left\langle v^2\right\rangle^{1/2}=(350 \pm 100)\kms$ for the {\tt 2D-ILC}
map.  One can see that the value we find
is consistent with the velocity dispersion
estimated through studies of the peculiar velocity field
\citep[e.g.,][]{Riess00,Turnbull12,mascott12,Carrick15}.

Here we need to remember that what we measured is the line-of-sight
velocity dispersion, which contains both the large-scale bulk flow, and the
small-scale velocity and intrinsic dispersion \citep[see, e.g.,][]{mascott14}.
The prediction for the \rms\ bulk flow, equation~(22) of \cite{planck2013-XIII}
(or equation~4 in \citealt{MaPan14}), is based on linear perturbation theory
for the $\Lambda$CDM model and works only for the large-scale bulk flows.
The small-scale motions and intrinsic dispersion are not fully predictable
from linear perturbation theory
because they depend on sub-Jeans scale structure evolution, which involves
nonlinear effects. However, this small-scale velocity and intrinsic dispersion
are nevertheless physical effects, which are non-negligible in
general \citep{Carrick15}. One should consider that
the line-of-sight velocity dispersion that we have measured is a
combination of two effects, namely large-scale bulk flows and small-scale intrinsic
dispersion, where the second component is generally non-negligible.

We estimate that the histogram of separation distances between all pairs of
cluster is peaked at $d \simeq 600\,{\rm Mpc}$.
Since the bulk flow contributes to the velocity dispersion measurement here,
then we can set an upper limit on the cosmic bulk flow on scales of
$600\,h^{-1}{\rm Mpc}$, $\langle v_{\rm bulk}^{2}\rangle^{1/2}<554\kms$
(95\,\% CL).
Such a constraint on large-scale bulk flows indicates that the Universe is
statistically homogeneous on scales of $600\,h^{-1}{\rm Mpc}$.
This is consistent with the limits obtained from
Type-Ia supernovae \citep{snbf2013}, the Spiral Field $I$-band
survey \citep{nusserdavis11,mascott12}, ROSAT galaxy clusters \citep{Mody12},
and the \Planck\ peculiar velocity study \citep{planck2013-XIII}.
However, it does not allow the very large ``dark flow'' claimed
in \cite{kashlinsky08,kashlinsky10,kashlinsky12} and \cite{Barandela15}. 
In addition to ruling out such models,
improved measurements of the velocity dispersion in the future have the
potential to set up interesting constraints on dark energy and modified
gravity \citep{Bhattacharya07,Bhattacharya08DE}.

\section{Conclusions}
\label{sec:conclusions}

The kinetic Sunyaev-Zeldovich effect gives anisotropic perturbations of the
CMB sky, particularly in the direction of clusters of galaxies.  Previous
studies have detected the kSZ effect through the pairwise momentum
estimator and temperature-velocity cross-correlation. In this paper, we have
detected the kSZ effect through a measurement of the temperature dispersion
and then we have interpreted this as a determination of the small-scale
velocity dispersion of cosmological structure.

To do this, we first selected two sets of \Planck\ foreground-cleaned maps.
One set contains four \Planck\ publicly available maps, namely \smica, \nilc,
\sevem, and \commander, each being produced using a different algorithm to
minimize foreground emission. The second set, is the \Planck\ {\tt 2D-ILC}
map, which nulls the tSZ component, while resulting in slightly larger residual
noise in the map. We then apply a matched-filter technique to the maps,
to suppress the primary CMB and instrumental noise.
We specifically consider the MCXC cluster sample.  Applying a Galactic
and point-source mask to the maps, results in $1526$ MCXC clusters
remaining unmasked.

We measured the distribution of the $\Delta T/T$ values for the $1526$ MCXC
cluster positions, and also at $1526$ randomly selected positions, to give a
quantification of the noise level. We found that the $1526$ true cluster
positions give extra variance to the distribution, and identify this as being
due to the kSZ temperature dispersion effect. We compare this signal to
results from 5000 random catalogues on the sky, each composed of $1526$ random
positions.  This extra dispersion signal is persistent in several tests that
we carry out.

We then construct estimators $\widehat{s^{2}}$ to quantify this effect. For
the \smica, \nilc, and \sevem\ maps, the significance of detection is stronger
than in the {\tt 2D-ILC} map, which is likely due to
the fact that the residual tSZ effect in the map is correlated with the kSZ
signal. In addition, an analytical estimate, supported by simulations, shows that about
$0.7\,\sigma$ of the temperature-dispersion effect comes from gravitational lensing. By subtracting
this effect, quoting the conservative result from {\tt 2D-ILC}, we obtain $\langle
s^{2} \rangle =(1.35 \pm 0.48)\times 10^{-11}$ (68\,\% CL), where $\langle
s^{2} \rangle =N_{\rm c}^{-1}\sum_{j}\left(\Delta T_{j}/T \right)^{2}$ ($N_{\rm c}=1526$).
This gives a detection of temperature dispersion at about the $2.8\,\sigma$
level. We obtain largely consistent results when we obtain results by weighting clusters
with their different observed properties.

We further estimate the optical depth of each cluster, and thereby convert our
(lensing-corrected) temperature dispersion measurement into a velocity
dispersion measurement,
obtaining $\langle v^{2}\rangle= (12.3 \pm 7.1) \times (100\,\kms)^{2}$ (68\,\%
CL) using a Gaussian approximation. The distribution has $P(v^{2}<0)=5.3$\,\%,
and the best-fit value is consistent with findings from large-scale structure
studies. This constraint implies that the Universe is statistically homogeneous
on scales of $600\,h^{-1}{\rm Mpc}$, with the bulk flow constrained to be $\langle v_{\rm bulk}^{2}\rangle^{1/2}<554\kms$ (95\,\% CL).

The measurement that we present here shows the promise of statistical kSZ
studies for constraining the growth of structure in the Universe.
To improve the results in the future, one needs to have better
component-separation algorithms to down-weight the residual noise contained in
the kSZ map, as well as having more sensitive and higher resolution CMB maps
for removing the tSZ signal.  One also needs larger
cluster catalogues, with the uncertainty scaling roughly as $1/\sqrt{N}$ if the
residual noise is Gaussian.

\begin{acknowledgements}
The Planck Collaboration acknowledges the support of: ESA; CNES, and
CNRS/INSU-IN2P3-INP (France); ASI, CNR, and INAF (Italy); NASA and DoE
(USA); STFC and UKSA (UK); CSIC, MINECO, JA, and RES (Spain); Tekes, AoF,
and CSC (Finland); DLR and MPG (Germany); CSA (Canada); DTU Space
(Denmark); SER/SSO (Switzerland); RCN (Norway); SFI (Ireland);
FCT/MCTES (Portugal); ERC and PRACE (EU). A description of the Planck
Collaboration and a list of its members, indicating which technical
or scientific activities they have been involved in, can be found at
\href{http://www.cosmos.esa.int/web/planck/planck-collaboration}{\texttt{http://www.cosmos.esa.int/web/planck/planck-collaboration}}.
This paper makes use of the {\tt HEALPix} software package.
\end{acknowledgements}

\bibliographystyle{aat}
\bibliography{pipkSZ,Planck_bib}

\appendix
\section{The statistics of {\boldmath $s^{2}$}}
\label{sec:AppA}
Let us first define
the $k$th moment of a distribution of a random variable $x$ to be
\begin{eqnarray}
\mu_{k}(x) \equiv E[x^{k}] =\frac{1}{N} \sum_{i} x^{k}_{i} .\label{eq:k-moment}
\end{eqnarray}
The $\widehat{s^{2}}$ estimator is defined in Eq.~(\ref{eq:def-s2}). Note that
the observed $\delta^{2}_{i}$ is always taken to be the value of kSZ on the
true cluster position, so there is no randomness in $\delta^{2}_{i}$. We also
define
\begin{eqnarray}
\overline{\delta^{2}} \equiv \frac{1}{N_{\rm c}}\sum \delta^{2}_{i}.
\end{eqnarray}
Therefore, the mean value of $\widehat{s^{2}}$ is
\begin{eqnarray}
E[s^{2}] &=& \overline{\delta^{2}}- \left(\frac{1}{N_{\rm c}}\sum_{i}E[n_{i}^{2}] \right) \nonumber \\
&=& \overline{\delta^{2}} -\mu_{2}(n), \label{eq:Es2}
\end{eqnarray}
while the variance of $\widehat{s^{2}}$ is
\begin{eqnarray}
V[s^{2}]=E[s^{4}]-\left(E[s^{2}]\right)^{2}.
\end{eqnarray}
Therefore, we first calculate
\begin{eqnarray}
s^{4} &=& \left[ \overline{\delta^{2}}- \left(\frac{1}{N_{\rm c}}\sum_{i}n_{i}^{2} \right)
 \right]^{2} \nonumber \\
&=& \left(\overline{\delta^{2}} \right)^{2}+\frac{1}{N^{2}_{\rm c}} \sum_{ij}n^{2}_{i}n^{2}_{j}-\frac{2}{N_{\rm c}}\overline{\delta^{2}} \left(\sum_{i} n^{2}_{i} \right) \nonumber \\
&=& \left(\overline{\delta^{2}} \right)^{2}+\frac{1}{N_{\rm c}} \sum_{i}n^{4}_{i} +\frac{1}{N^{2}_{\rm c}}\sum_{i,j \,(i \neq j)}n^{2}_{i}n^{2}_{j} \nonumber \\
&-& \frac{2}{N_{\rm c}}\overline{\delta^{2}} \left(\sum_{i} n^{2}_{i} \right).
\end{eqnarray}
And hence,
\begin{eqnarray}
E[s^{4}] &=& \left(\overline{\delta^{2}} \right)^{2}+\frac{1}{N_{\rm c}} \mu_{4}(n) +\frac{N_{\rm c}-1}{N_{\rm c}}\left(\mu_{2}(n) \right)^{2}  \nonumber \\
&-& 2\overline{\delta^{2}} \mu_{2}(n),
\end{eqnarray}
where in the above derivation we have assumed that the residual noise samples
in two different pixels are uncorrelated, i.e., $\langle n_{i}n_{j} \rangle=0$.
Finally,
\begin{eqnarray}
V[s^{2}]=\frac{1}{N_{\rm c}}\left(\mu_{4}(n)-\mu^{2}_{2}(n) \right). \label{eq:Vs2}
\end{eqnarray}

\section{Converting {\boldmath $M_{500}$} to {\boldmath $M_{\rm vir}$}}
\label{sec:m500-mvir}

For each cluster, $M_{500}$ is defined as the mass within the radius of
$R_{500}$, in which its average density is $500$ times the critical density of
the Universe,
\begin{eqnarray}
M_{500}=\frac{4\pi}{3}\left[500 \rho_{\rm c}(z)
\right]r^{3}_{500}, \label{eq:M500-1}
\end{eqnarray}
where
\begin{eqnarray}
\rho_{\rm c}(z)=2.77h^2E^2(z) \times 10^{11}{\rm M}_{\odot}{\rm Mpc}^{-3},
\end{eqnarray}
and $E^{2}(z)=\Omega_{\rm m}(1+z)^{3}+\Omega_{\rm \Lambda}$.
The quantity
$M_{\rm vir}$ is calculated via Eqs.~(\ref{eq:M-vir}) and (\ref{eq:Delta-z}),
and the relationship between $M_{500}$ and $M_{\rm vir}$ is \citep{Mody12}
\begin{eqnarray}
\frac{M_{500}}{M_{\rm vir}}=\frac{m(cr_{500}/r_{\rm vir})}{m(c)}, \label{eq:M-ratio}
\end{eqnarray}
where $c$ is the concentration parameter (Eq.~\eqref{eq:c-para}) and
$m(x)=\ln(1+x)-x/(1+x)$.
Given redshift $z$ and mass $M_{500}$, we can thus determine $r_{500}$ through
Eq.~(\ref{eq:M500-1}). If we substitute $M_{500}$, $z$, and $r_{500}$ into
Eq.~(\ref{eq:M-ratio}), this becomes an algebraic equation for $M_{\rm vir}$.
This is because $r_{\rm vir}$ can be determined from $M_{\rm vir}$ through
Eqs.~(\ref{eq:M-vir}) and (\ref{eq:Delta-z}), and $c$ is related to
$M_{\rm vir}$ through Eq.~(\ref{eq:c-para}). Therefore, we can iteratively
solve for $M_{\rm vir}$, given the values of $M_{500}$ and $z$.

\section{Converting \textit{P} values into S/N ratios}
\label{sec:P-SN}

Since the distribution of weighted $s^{2}$ has a longer tail than a Gaussian
distribution, instead of calculating the ratio between mean and \rms\ values
of the distribution, we calculate the $p$-values, and list them in the third
column of Table~\ref{tab:stats-weights}. We now convert them into
S/N in the following way.  Suppose the variable $x$ satisfies a Gaussian
distribution, the normalized distribution is
$L(x)=(1/\sqrt{2\pi}\sigma)\exp(-(x-\mu)^{2}/2\sigma^{2})$. Then the cumulative
probability to find $x<0$ is
\begin{eqnarray}
\epsilon=\int^{0}_{-\infty}P(x)\der x =\frac{1}{2}{\rm Erfc}\left(\frac{\mu}{\sqrt{2}\sigma} \right),
\end{eqnarray}
where
\begin{eqnarray}
{\rm Erfc}(x)=\frac{2}{\sqrt{\pi}}\int^{\infty}_{x}{\rm e}^{-t^{2}}\der t,
\end{eqnarray}
is the complimentary error function.

Therefore the equivalent S/N given the value of $P(s^{2}_{\rm w}<0)$ is
\begin{eqnarray}
{\rm S/N}=\sqrt{2}\left\{{\rm Erfc}^{-1}\left[2P(s^{2}_{\rm w}<0) \right] \right\}. \label{eq:SN-P}
\end{eqnarray}
We use Eq.~(\ref{eq:SN-P}) to obtain the fourth column of Table~\ref{tab:stats-weights}.

\raggedright
\end{document}